\documentclass[12pt]{article}

\usepackage[english]{babel}
\usepackage[T1]{fontenc}
\usepackage[utf8]{inputenc}
\usepackage{epsf,amssymb,amsmath,array,hhline, dsfont}
\usepackage{latexsym}
\usepackage{graphicx}
\usepackage[final]{pdfpages}
\usepackage{verbatim}
\usepackage{stmaryrd}

\title{Scattering zippers and their spectral theory}

\author{Laurent Marin, Hermann Schulz-Baldes
\\
\\
{\small Department Mathematik, Universit\"at Erlangen-N\"urnberg, Germany}
}

\date{ }

\newtheorem{theo}{Theorem}
\newtheorem{defini}{Definition}
\newtheorem{proposi}{Proposition}
\newtheorem{lemma}{Lemma}
\newtheorem{coro}{Corollary}

\newcommand{\CM}{{\mathbb C}}

\newcommand{\SM}{{\mathbb S}}
\newcommand{\TM}{{\mathbb T}}
\newcommand{\ZM}{{\mathbb Z}}
\newcommand{\LM}{{\mathbb L}}

\newcommand{\HM}{{\mathbb H}}
\newcommand{\DM}{{\mathbb D}}
\newcommand{\GM}{{\mathbb G}}

\newcommand{\Pp}{{\cal P}}

\newcommand{\Ff}{{\cal F}}

\newcommand{\WQ}{\widetilde{\mathcal Q}}

\newcommand{\Tt}{{\cal T}}

\newcommand{\Mm}{{\cal M}}
\newcommand{\Cc}{{\cal C}}
\newcommand{\CMV}{{\mathbb U}}
\newcommand{\CMVL}{{\mathbb V}}
\newcommand{\CMVR}{{\mathbb W}}
\newcommand{\Jj}{{\cal J}}

\newcommand{\Ll}{{\cal L}}
\newcommand{\Qq}{{\cal Q}}

\newcommand{\one}{{\bf 1}}

\newcommand{\per}{{\mbox{\rm\tiny per}}}
\newcommand{\inv}{{\mbox{\rm\tiny inv}}}

\newcommand{\Weyl}{\mathfrak{W}}

\begin{document}

\maketitle

\begin{abstract}
A scattering zipper is a system obtained by concatenation of scattering events with equal even number of incoming and out going channels. The associated scattering zipper operator is the unitary equivalent of Jacobi matrices with matrix entries and generalizes Blatter-Browne and Chalker-Coddington models and CMV matrices. Weyl discs are analyzed and used to prove a bijection between the set of semi-infinite scattering zipper operators and matrix valued probability measures on the unit circle. Sturm-Liouville oscillation theory is developed as a tool to calculate the spectra of finite and periodic scattering zipper operators.
\end{abstract}

\vspace{.5cm}

\section{Scattering zippers}

A scattering zipper describes consecutive scattering events with a fixed number $2L$ of incoming and out-going channels each. It is specified by a sequence $(S_n)_{n=2,\ldots,N}$ of unitary scattering matrices $S_n$ in the unitary group $\mbox{U}(2L)$ as well as two unitaries $U,V\in\mbox{U}(L)$ modeling the boundary scattering. The size $N$ of the system is supposed to be either even or infinite. Then the scattering zipper operator acting on $\ell^2(\{1,\ldots,N\},\CM^L)$ is defined as 
$$
\CMV_N\;=\;\CMVL_N\,\CMVR_N\;,
$$
where the two unitaries $\CMVL_N$ and $\CMVR_N$ are given by 
$$
\CMVL_N
\;=\; 
\begin{pmatrix}
S_2 & & & & \\
 & S_4 & & & \\
& & \ddots & & \\
& & & \ddots & \\
& & & & S_{N}
\end{pmatrix}                                                                                            
\;,
\qquad
\CMVR_N
\;=\;
\begin{pmatrix}
U & & & & \\
& S_3  & & & \\
 & & \ddots  & & \\
& & & S_{N-1}  & \\
& & & & V
\end{pmatrix}                                                                                            
\;.
$$
The main hypothesis on each of the scattering matrices $S_n$ is that its upper right entry of size $L\times L$ is invertible. In the notation of Section~\ref{sec-prel} below, $S_n$ is in a subset $\mbox{\rm U}(2L)_\inv$ of the unitary group. This condition assures that the scattering is effective so that $\CMV_N$ does not decouple into a direct sum of two or more parts. The terminology {\it scattering zipper} is best understood by looking at Figure~1 illustrating the model. It shows the first scattering events of a semi-infinite model $N=\infty$ for which we also drop the indices on $\CMV$, $\CMVL$ and $\CMVR$. It is also possible to consider periodic scattering zippers, see Section~\ref{sec-periodic} below for finite operators and Section~\ref{sec-periodic-infinite} for infinite ones. Furthermore, by placing either of the boundary conditions $U$ and $V$ into $\CMVL_N$, one can consider the case of odd $N$, but we refrain from doing so.

\begin{figure}
\begin{center}
\includegraphics{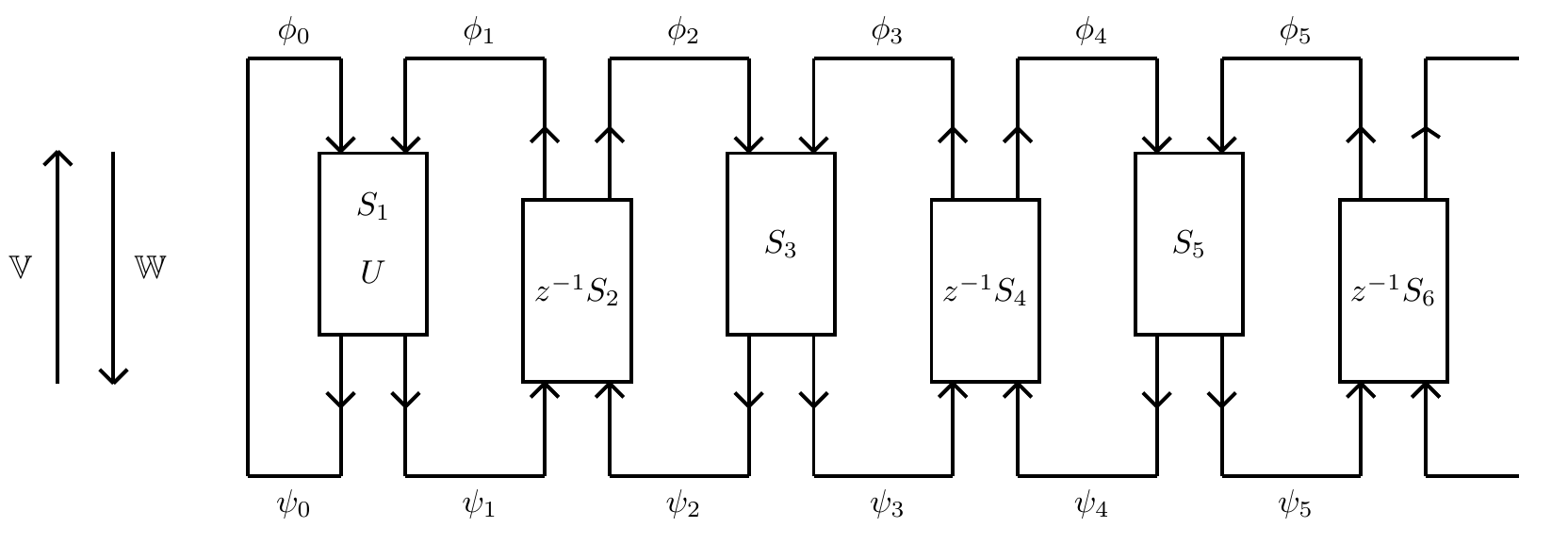}
\caption{\it Illustration of the first scattering events as well as the boundary scattering. The wave functions $\phi^z=(\phi^z_n)_{n\geq 1}$ and $\psi^z=(\psi^z_n)_{n\geq 1}$ satisfy $\CMV\phi^z=z\phi^z$, which is equivalent to $\phi^z=z^{-1}\CMVL\psi^z$ and $\psi^z=\CMVR\phi^z$. The picture corresponds to {\rm Proposition~\ref{prop-solutions}} below. The boundary conditions $\phi^z_0$ and $\psi^z_0$ as well as $S_1$ can be introduced for convenience, but one can also just keep $U$, {\it c.f.}  {\rm Section~\ref{sec-transfer}}.
}
\end{center}
\end{figure}

\vspace{.2cm}

The main message of this paper is that scattering zippers are the unitary analogs of Jacobi matrices with matrix entries. Here are the structural results supporting this claim:

\begin{itemize}

\item The matrix $\CMV_N$ is not tridiagonal, but five-diagonal. Nevertheless, solutions of the associated eigenvalue equation can be calculated by transfer matrices having the same symmetries as in the Jacobi case, but no further restrictions (see Section~\ref{sec-transfer}).

\item The matrix element of the resolvent of $\CMV_N$ corresponding to the left boundary site $1$ has a simple expression in terms of the entries of the transfer matrix, namely it is given by a M\"obius transformation of the other boundary condition  (Theorem~\ref{theo-Green}).

\item These matrix elements of the resolvent lie on a Weyl surface which is a matrix ball (Theorems~\ref{theo-Weyl1} and \ref{theo-Weylsurf}).

\item All semi-infinite scattering zipper operators $\CMV$ with fixed boundary condition $U$ are in the limit point case. Fixing an appropriate gauge for each $S_n$, the semi-infinite scattering zipper operators are in bijection with their spectral measures which are all matrix-valued probability measures on the unit circle (Theorem~\ref{theo-bijection}).

\item The eigenvalues of the finite scattering zipper operators can be calculated using matrix Pr\"ufer phases by Sturm-Liouville type oscillation theory (Theorems~\ref{theo-oscillation} and \ref{theo-osci2}). This is also an efficient tool to calculate the spectrum of infinite periodic scattering zippers (Theorem~\ref{theo-oscillationper}).

\end{itemize}

For Jacobi matrices with matrix entries, it is well-known how to use transfer matrices, and how to calculate the resolvents ({\it e.g.} \cite{SB1}, but this is classical). Weyl surfaces are also known for Jacobi matrices (see \cite{SB3} which also contains earlier references). The reason why there is a simple bijection in Theorem~\ref{theo-bijection} is that all semi-infinite scattering zipper operators are in the Weyl limit point case, namely the Weyl surfaces shrink to one point in the large $N$ limit. This is in strong contrast with Jacobi matrices where there are limit discs which lead to all the issues related to the moment problem (see {\it e.g.}  the beautiful paper by Simon \cite{Sim1}), nevertheless  there are close connections between probability measures on the real line and Jacobi matrices. Finally, oscillation theory of Jacobi matrices with matrix entries was developed in \cite{SB1,SB2}, but again the scalar case is well-known.

\vspace{.2cm}

Now let us present our personal and without doubt very restricted view on connections of this work to the literature.  First of all, the scattering zipper itself is a generalization of three well-known models, the Blatter-Browne model \cite{BB} and the Chalker-Coddington network model \cite{CC} of the solid state physics community as well as the CMV matrices \cite{CMV} of the mathematical literature. The Blatter-Browne model is scalar, namely $L=1$. In the Chalker-Coddington model, the aim is to model a higher dimensional lattice of scattering events. In terms of the scattering zipper this means that there is supplementary structure in each of the scattering matrices $S_n$ and that they are infinite-dimensional and very sparsely filled. Motivated by applications to the quantum Hall transitions, the main focus in the Chalker-Coddington model has been on random scattering events. First rigorous works on the analysis of the Blatter-Brown model and the Chalker-Coddington model on a strip have appeared \cite{BHJ,ABJ}. On the other hand, the CMV matrices in its matricial version \cite{DPS} only consider scattering blocks of the type
\begin{equation}
\label{eq-CMV}
S_n\; =\; 
\begin{pmatrix} \alpha_n & (\one-\alpha_n\alpha_n^*)^{\frac{1}{2}} 
\\ 
(\one-\alpha_n^*\alpha_n)^{\frac{1}{2}} & -\alpha_n^* 
\end{pmatrix}
\;,
\end{equation}
where the $\alpha_n$ verify $\alpha_n^*\alpha_n <\one$ which is equivalent to $\|\alpha_n\|<1$ (here and below we always use the operator norm). These $\alpha_n$ are called the Verblunsky coefficients \cite{DPS}. There is a huge literature on CMV matrices (see \cite{Sim,DPS} for a long list of references). They form a subclass of the scattering zipper models considered here. Of course, the transfer matrix techniques also apply and have very efficiently been used in most works on the subject. However, the Weyl discs at finite $N$ as presented below seem to have been studied only in the scalar case \cite{GN}. Nevertheless, it was possible to prove in \cite{CGZ,DPS} that every sequence of Verblunsky coefficients corresponds to a unique matrix-valued measure on the unit circle (Verblunsky's theorem). Theorem~\ref{theo-bijection} extends this theorem in that it exhibits a bijection between all probability measures on the unit circle and the semi-infinite scattering zippers with fixed boundary condition $U$. Finally, it seems that oscillation theory for CMV matrices was only developed in the scalar case $L=1$ \cite[Theorem 8.3.3.]{Sim}.

\vspace{.2cm}

\noindent {\bf Acknowlegements:} H.~S.-B. wants to thank Mihai Stoiciu for introducing him to the world of CMV matrices and for a number of discussions about oscillation theory at a very early stage of this work. He also let us know about the reference \cite{BB}. We also acknowlege financial support of the DFG.

\section{Preliminaries on scattering matrices and transfer matrices}
\label{sec-prel}

As explained in the introduction, the following subset of the even-dimensional unitary group will play a role:
\begin{equation}
\label{eq-U(2L)subset}
\mbox{U}(2L)_\inv
\; =\;
\left\{\left. S = \begin{pmatrix} \alpha & \beta \\ \gamma & \delta \end{pmatrix} \in \mbox{U}(2L), \;\right|\; \beta\;
\mbox{\rm is an invertible $L\times L$ matrix}\; 
\right\}
\;.
\end{equation}

\begin{proposi}
In {\rm \eqref{eq-U(2L)subset}}, equivalent to the condition that $\beta$ is invertible is either the invertibility of $\gamma$ or the condition $\alpha^*\alpha<\one$ or the condition $\delta^*\delta<\one$.
Furthermore, one has the representation
\begin{equation}
\label{eq-CMVrep}
\mbox{\rm U}(2L)_\inv
\; =\;
\left\{\left. S(\alpha,U,V) \in \mbox{\rm U}(2L) \;\right|\; \alpha^*\alpha<\one \;\mbox{\rm and }U,V\in\mbox{\rm U}(L) 
\right\}
\;,
\end{equation}
where
\begin{equation}
\label{eq-Smatdef}
S(\alpha,U,V)\; =\; 
\begin{pmatrix} \alpha & (\one-\alpha\alpha^*)^{\frac{1}{2}}U 
\\ 
V(\one-\alpha^*\alpha)^{\frac{1}{2}} & -V\alpha^*U 
\end{pmatrix}
\;.
\end{equation}
\end{proposi}

\noindent {\bf Proof.} The equations $S^*S = \one = SS^*$ give
\begin{eqnarray}
\alpha^*\alpha + \gamma^*\gamma = \one \;,\qquad \; \delta^*\delta + \beta^*\beta = \one\;,
\qquad 
\delta^*\gamma+ \beta^*\alpha =0 \;,
\label{eq-relation1}
\\
\alpha\alpha^* + \beta\beta^* = \one\;, 
\qquad \delta\delta^* + \gamma\gamma^* = \one\;,\qquad
\gamma\alpha^* + \delta\beta^* = 0 \;.
\label{eq-relation2}
\end{eqnarray}
From three of these identities the first claims can be deduced immediately. For \eqref{eq-CMVrep}, let us first of all note that $\beta\beta^* =\one-\alpha\alpha^*$ implies that $\beta^*$ has a unique polar decomposition $\beta^*=U^*(\one-\alpha\alpha^*)^{\frac{1}{2}}$ with some unitary $U$. Similarly, $\gamma^*\gamma = \one-\alpha^*\alpha$ shows that $\gamma=V(\one-\alpha\alpha^*)^{\frac{1}{2}}$ for some unitary $V$. But then $\delta=-\gamma\alpha^*(\beta^*)^{-1}=-V(\one-\alpha^*\alpha)^{\frac{1}{2}}\alpha^*(\one-\alpha\alpha^*)^{-\frac{1}{2}}U=-V\alpha^*U$.
\hfill $\Box$

\vspace{0.2cm}

Recall that the Lorentz group $\mbox{U}(L,L)$ of signature $(L,L)$ is defined to be the set of $2L\times 2L$ matrices conserving the form
\begin{equation}
\label{eq-Ldef}
\Ll
\;=\;
\begin{pmatrix}
 \one & 0
\\
0 & -\one
\end{pmatrix}
\;.
\end{equation}
The following well-known result on the passage from scattering matrices to transfer matrices is illustrated in Figure~2.

\begin{proposi}
\label{prop-Stransferequiv}
The formula
$$
\varphi \left(\begin{pmatrix}  \alpha & \beta \\ \gamma & \delta \end{pmatrix}\right) 
\;=\; 
\begin{pmatrix} \gamma-\delta\beta^{-1}\alpha & \delta\beta^{-1} \\ -\beta^{-1} \alpha & \beta^{-1} \end{pmatrix}
$$
defines a bijection from $\mbox{\rm U}(2L)_\inv$ onto $\mbox{\rm U}(L,L)$. For any vectors $\phi,\phi',\psi,\psi'\in\CM^{L}$, one has the equivalence
\begin{equation}
\label{eq-Stransferequiv}
S \begin{pmatrix} \psi \\ \psi' \end{pmatrix}
\;=\;
\begin{pmatrix} \phi \\ \phi' \end{pmatrix}
\qquad
\Longleftrightarrow
\qquad
\varphi(S) \begin{pmatrix} \psi \\ \phi \end{pmatrix}
\;=\;
\begin{pmatrix} \phi' \\ \psi' \end{pmatrix}
\;.
\end{equation}
Moreover, if $\xi,\xi'\in\CM^L$ form an inhomogeneity, then
\begin{equation}
\label{eq-Stransferequiv2}
S \begin{pmatrix} \psi \\ \psi' \end{pmatrix}
\;=\;
\begin{pmatrix} \phi \\ \phi' \end{pmatrix}
\,+\,
\begin{pmatrix} \xi \\ \xi' \end{pmatrix}
\qquad
\Longleftrightarrow
\qquad
\varphi(S) \begin{pmatrix} \psi \\ \phi \end{pmatrix}
\;=\;
\begin{pmatrix} \phi' \\ \psi' \end{pmatrix}
\,+\,
\begin{pmatrix} -\delta\beta^{-1} & \one \\ -\beta^{-1} & 0 \end{pmatrix}
\begin{pmatrix} \xi \\ \xi' \end{pmatrix}
\;.
\end{equation}
\end{proposi}

\noindent {\bf Proof.} First of all, note that $\varphi$ is well-defined because $\beta$ is invertible. Next one readily checks that $\varphi(S)^*\Ll\varphi(S)=\Ll$ by using the relations \eqref{eq-relation1} and \eqref{eq-relation2}. Moreover, the inverse of $\varphi$ is given by
$$
\varphi^{-1}\left(\begin{pmatrix}  A & B \\ C & D \end{pmatrix}\right) 
\;=\; 
\begin{pmatrix} -D^{-1}C & D^{-1} \\ A-BD^{-1}C & BD^{-1} \end{pmatrix}
\;.
$$ 
Next let us check \eqref{eq-Stransferequiv2}, which generalizes \eqref{eq-Stransferequiv}. The upper equation on the l.h.s. is $\alpha\psi+\beta\psi'=\phi+\xi$ which can be rewritten as
$$
-\beta^{-1}\alpha\psi+\beta^{-1}\phi\;=\;\psi'-\beta^{-1}\xi
\;.
$$
This is already the lower equation of the r.h.s.. Solving it for $\psi'$ and replacing in the lower equation of the l.h.s. gives the upper equation on the r.h.s..
\hfill $\Box$

\vspace{.2cm}

\begin{figure}
\begin{center}
\includegraphics{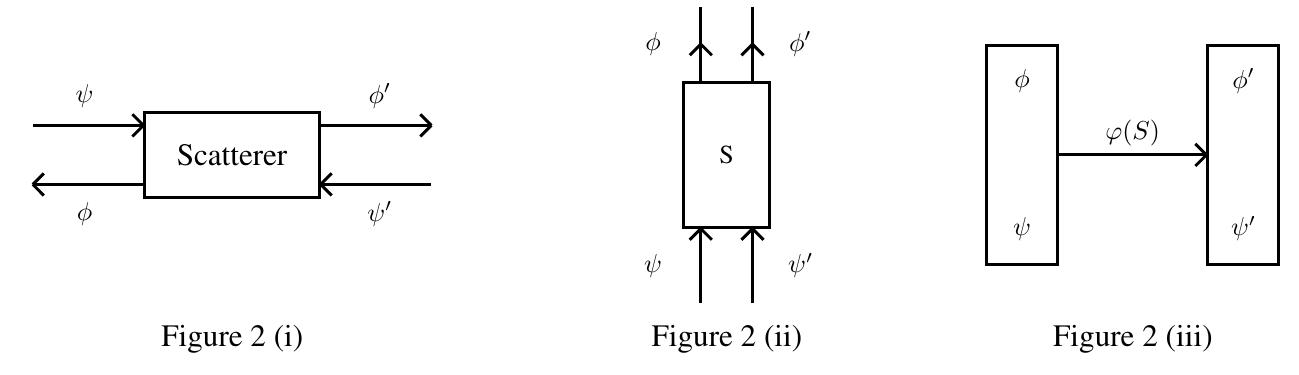}
\caption{\it Illustration of the scattering event described in {\rm Proposition~\ref{prop-Stransferequiv}}. {\rm Figure~2(i)} is the usual graphical representation of a scattering event with incoming amplitudes $\psi$ and $\psi'$ and outgoing amplitudes $\phi$ and $\phi'$. {\rm Figures~2(ii)} and {\rm 2(iii)} corresponding to the left and right hand side of equation~{\rm \eqref{eq-Stransferequiv}} show different representations of the same event as they are used as building blocks in {\rm Figure 1} and {\rm 3} respectively.  
}
\end{center}
\end{figure}

Let us conclude this section with a few comments. First of all, one readily checks
$$
\varphi\bigl(S(\alpha,U,V)\bigr)
\;=\;
\begin{pmatrix} V & 0 \\ 0 & U^* 
\end{pmatrix}
\,
\begin{pmatrix} (\one-\alpha^*\alpha)^{-\frac{1}{2}}  & (\one-\alpha^*\alpha)^{-\frac{1}{2}}\alpha^*
\\ 
\alpha(\one-\alpha^*\alpha)^{-\frac{1}{2}} & (\one-\alpha\alpha^*)^{-\frac{1}{2}}
\end{pmatrix}
\;.
$$
Comparing with \eqref{eq-CMV}, one way to characterize CMV matrices is therefore to say that its individual  scattering events give rise to self-adjoint transfer matrices. Of course, this does not imply that products of such transfer matrices have the same property. This means that CMV matrices are not specified by some symmetry given by a combination of time-reversal, particle-hole or sublattice symmetry, and it is not clear to us whether there is a deeper reason to consider scattering events of the type \cite{CMV}. On the other hand, it is possible to implement all the above symmetries also in scattering zippers, just as for Jacobi matrices with matrix entries (see \cite{SB1} where only even and odd time-reversal symmetry is dealt with). Finally, let us briefly discuss degenerate scattering events which don't mix all incoming and outgoing amplitudes. There are many possibilities to do this, but only two are relevant for the boundary conditions in the next section. In one $\phi$ only depends on $\psi$ (left and right in Figure~1(i) decoupled), in the other only on $\psi'$ (top and bottom in Figure~1(i) decoupled). Let us focus on the latter. Then there are two unitaries $U$ and $V$ such that $\phi=U\psi'$ and $\phi'=V\psi$. In this case, $S$ and $\varphi(S)$ are given by
$$
S\;=\;
\begin{pmatrix}
0 & U \\ V & 0 
\end{pmatrix}
\;,
\qquad
\varphi(S)
\;=\;
\begin{pmatrix}
V & 0 \\ 0 & U^* 
\end{pmatrix}
\;.
$$

\section{Solutions and transfer matrices}
\label{sec-transfer}

In this section, the formal solutions $\phi^z = (\phi^z_n)_{n\geq 1}$ for the eigenvalue equation $\CMV\phi^z=z\phi^z$ at $z\in\CM$ will be constructed. Here all the $\phi^z_n$ are $L\times L$ matrices and the index $n$ runs to infinity and it is formal in the sense that $\phi^z$ is typically not square integrable. The construction of $\phi^z$ is done such that the left boundary condition (at site $1$) is satisfied. For finite $N$, the solution $\phi^z$ in general does not satisfy the right boundary condition, namely $(\CMV_N\phi^z)_N\not =z\phi^z_N$.

\begin{proposi}
\label{prop-solutions}
Let $\phi^z_1=\one$. Then the following assertions are equivalent:

\vspace{.1cm}

\noindent {\rm (i)} $(\CMV\phi^z)_n = z\phi^z_n$ and $(\CMVR \phi^z)_n =  \psi^z_n$ for $ n \geq 1$.

\vspace{.1cm}

\noindent {\rm (ii)} $(\CMVR \phi^z)_n =  \psi^z_n$ and $(\CMVL \psi^z)_n = z \phi^z_n$ for $ n \geq 1$.

\vspace{.1cm}

\noindent {\rm (iii)} For any $k\geq 1$,
$$
\begin{pmatrix}  \phi^z_{2k} \\ \psi^z_{2k}\end{pmatrix}\; =\; \mathcal{T}^z_{2k} 
\begin{pmatrix} \psi^z_{2k-1} \\ \phi^z_{2k-1} \end{pmatrix}
\;,
\qquad
\begin{pmatrix}  \psi^z_{2k+1} \\ \phi^z_{2k+1}\end{pmatrix}\; =\; \mathcal{T}^z_{2k+1} 
\begin{pmatrix} \phi^z_{2k} \\ \psi^z_{2k} \end{pmatrix}
\;,
$$

where the transfer matrices and initial condition are
$$
\Tt^z_{2k}\;=\;\varphi(z^{-1} S_{2k})\;,
\qquad
\Tt^z_{2k+1}\;=\;\varphi(S_{2k+1})\;,
\qquad
\begin{pmatrix} \psi^z_1 \\ \phi^z_1 \end{pmatrix} 
\;=\; \begin{pmatrix} U \\ \one  \end{pmatrix}
\;.
$$
\end{proposi} 

\begin{figure}
\begin{center}
\includegraphics{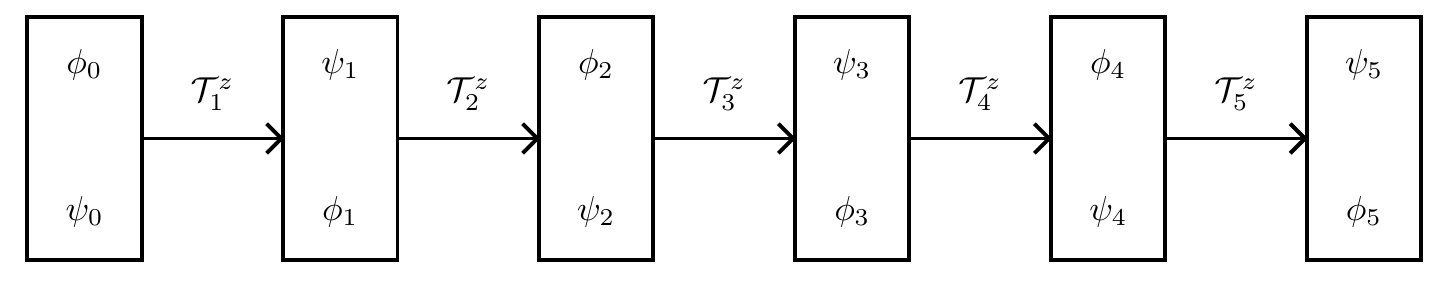}
\caption{\it The scattering zipper of {\rm Figure 1} after having transformed each scattering
event by $\varphi$ into a transfer matrix multiplication as done in {\rm Proposition~\ref{prop-solutions}}.
}
\end{center}
\end{figure}

\noindent {\bf Remark}
The application $\varphi$ is used here even if $z$ is not on the unit circle so that $z^{-1}S_{2l}$ is not unitary. Of course, $\Tt^z_{2k}$ is in the group $\mbox{U}(L,L)$ only if $z$ is on the unit circle. Also let us point out that $\Tt^z_{2k+1}$ is actually independent of $z$. Moreover, one could also use the initial condition 
$$
\begin{pmatrix} \phi^z_0 \\ \psi^z_0 \end{pmatrix} 
\;=\; \begin{pmatrix} \one \\ \one  \end{pmatrix}
\;,
$$
and add in (iii) one more equation, namely
$$
\begin{pmatrix}  \psi^z_{1} \\ \phi^z_{1}\end{pmatrix}\; =\; \mathcal{T}^z_{1} 
\begin{pmatrix} \phi^z_{0} \\ \psi^z_{0} \end{pmatrix}
\;,
\qquad
\Tt^z_1\;=\;
\varphi(S_1)\;,
\qquad
S_1
\;=\;
\begin{pmatrix}
0 & \one \\ U & 0 
\end{pmatrix}
\;.
$$
This results from the discussion at the end of Section~\ref{sec-prel}.
\hfill $\diamond$


\vspace{.2cm}

\noindent {\bf Proof} of Proposition~\ref{prop-solutions}: The equivalence of (i) and (ii) results immediately from $\CMV=\CMVL\CMVR$. The equivalence of (ii) and (iii) can be checked by successive application of the identity \eqref{eq-Stransferequiv} in Proposition~\ref{prop-Stransferequiv}. More precisely, the two equations of (ii) mean that for any $k \geq 1$
$$
S_{2k} \begin{pmatrix}    \psi^z_{2k-1}\\ \psi^z_{2k}\end{pmatrix}\; =\; z 
\begin{pmatrix} \phi^z_{2k-1} \\ \phi^z_{2k} \end{pmatrix}
\;,
\qquad
S_{2k+1} \begin{pmatrix} \phi^z_{2k} \\ \phi^z_{2k+1} \end{pmatrix}
\; =\; 
 \begin{pmatrix}    \psi^z_{2k}\\ \psi^z_{2k+1}\end{pmatrix}
\;.
$$
Applying the transformation $\varphi$ from Proposition~\ref{prop-Stransferequiv} to both of these equations than shows the equivalence with (iii).
\hfill $\Box$

\vspace{.2cm}

As usual, the transfer matrices can be nicely iterated. Let us set for $n\geq k$
$$
\Tt^z(n,k)
\;=\;
\Tt^z_n\cdots\Tt^z_{k+1}
\;,
$$
and in order to nicely write the solutions, let us also introduce the notations
\begin{equation}
\label{eq-Phidef}
\Phi_{2k+1}^z
\;=\;
\begin{pmatrix} \psi^z_{2k+1} \\ \phi^z_{2k+1} \end{pmatrix}
\;,
\qquad
\Phi_{2k}^z
\;=\;
\begin{pmatrix} \phi^z_{2k} \\ \psi^z_{2k} \end{pmatrix}
\;.
\end{equation}
Then the solutions are simply given by
\begin{equation}
\label{eq-recurrence}
\Phi_{n}^z
\;=\;
\Tt^z_{n}\,\Phi_{n-1}^z
\;=\;
\Tt^z(n,0)
\,\Phi_{0}^z
\;,
\qquad
\Phi_{0}^z\;=\;
\begin{pmatrix} \one \\ \one  \end{pmatrix}\;.
\end{equation}
%

\section{Resolvents}
\label{eq-resolvent}

As a preparation for the Weyl theory, this section calculates the entries of the resolvents of $\CMV_N(V)$ corresponding to the site $1$ in terms of the entries of the transfer matrix from $0$ to $N$:
$$ \Tt^z (N,0) \; = \;
\begin{pmatrix} A_N^z & B_N^z \\ C_N^z & D_N^z \end{pmatrix} 
\;.
$$ 
The first object of study is the Green matrix
$$
G^z_N(V)
\;=\;
\pi_1^*(\CMV_N(V)-z)^{-1} \pi_1
\;,
$$
where $\pi_n:\CM^L\to \ell^2(\{1,\ldots,N\},\CM^L)$ is the partial isometry onto the $n$th site and $z$ is in the open unit disc $\DM$. For self-adjoint matrices, the Green matrix has a Herglotz property, which it is laking in the present situation. Actually, for the unitary operator $\CMV_N(V)$ it is more natural to consider
$$
F^z_N(V)
\;=\;
\imath\;
\pi_1^*(\CMV_N(V)-z)^{-1} (\CMV_N(V)+z)\pi_1
\;.
$$
One readily checks that this analytic function $z\in\DM\mapsto F^z_N(V)\in\mbox{\rm Mat}(L,\CM)$ has a positive imaginary part $\imath(F(z)^*-F(z))>0$. It also satisfies $F(0)=\imath\,\one$ and is thus a so-called  Caratheodory function (up to the factor $\imath$). If one also takes the Cayley transform of the domain $\DM$ to the upper half-plane $\HM$, one has again a Herglotz function. A little more information on this is resembled in Appendix~B.
 
\begin{theo}
\label{theo-Green} 
The matrix $C^z_N - V A^z_N$ is invertible and
$$
E_N^z(V)\;=\; (C^z_N -VA^z_N)^{-1} (VB^z_N -  D^z_N)\;,
$$ 
lies in the Siegel disc $\DM_L=\{Z\in\mbox{\rm Mat}(L,\CM)\,|\,Z^*Z<\one\}$. Moreover, the resolvent matrix $F^z_N(V)$ is given by
\begin{equation}
F^z_N(V)
\; = \; 
\frac{1}{\imath}\;(E^z_N(V)+\one)\,
(E^z_N(V)-\one)^{-1} 
\;,
\label{eq-resol1}
\end{equation}
and the Green matrix by 
\begin{equation}
G^z_N(V)
\; = \;
\frac{1}{z}\;
E^z_N(V)\;
(\one -  E^z_N(V))^{-1} \;.  
\label{eq-Green1}
\end{equation}
\end{theo}

Let us note that the formulas for $E_N^z(V)$ and $F^z_N(V)$ can also be written using the (inverse) matrix M\"obius transformation (see Appendix A):
\begin{equation}
\label{eq-EFmoeb}
E_N^z(V) 
\;=\; V^* : \Tt^z (N,0)
\;,
\qquad
F^z_N(V)
\; = \;
\Cc^*\cdot
E^z_N(V) 
\;,
\end{equation}
where the Cayley transformation is the $2L\times 2L$ unitary matrix defined by
$$
\Cc
\;=\;
\frac{1}{\sqrt{2}}\;
\begin{pmatrix} \one & -\imath\,\one \\ \one & \imath\,\one \end{pmatrix}
\;.
$$
Based on \eqref{eq-EFmoeb} and the results of  Appendix~A, one obtains further identities, {\it e.g.}
\begin{equation}
\label{eq-EFmoeb2}
E_N^z(V) 
\;=\; \Tt^z (N,0)^{-1}\cdot V^* 
\;,
\qquad
F^z_N(V)
\; = \;
\bigl(\Cc^*\,\Tt^z (N,0)^{-1}\bigr)\cdot V^* 
\;.
\end{equation}
For the proof of Theorem~\ref{theo-Green}, one needs a number of lemmata which will also be useful later on for other purposes.

\begin{lemma}
\label{lem-lemmaP}
Let $\Tt^z=\varphi(z^{-1}S)$ for some $S\in\mbox{\rm U}_\inv(2L)$ and $z\in\overline{\DM}$, and set
$$
\mathcal{T}^z \; = \;
\begin{pmatrix}
z^{-1} A & B \\
C & zD
\end{pmatrix}
\;,
\qquad
\mathcal{P}^z \;=\; \begin{pmatrix}
\left( |z|^{-2}-1 \right) A^*A & \left(({\overline{z}})^{-1} -z \right) A^*B \\
\left( z^{-1} -\overline{z} \right) B^*A & \left( 1- |z|^2 \right) (B^*B+\one)
\end{pmatrix}
\;.
$$ 
Then $\mathcal{P}^z\geq \frac{1-|z|^{2}}{2}$ and
\begin{equation}
\label{eq-Prel}
(\mathcal{T}^z)^* \,\Ll\, \mathcal{T}^z 
\;= \;
\Ll + \mathcal{P}^z
\;.
\end{equation}
Moreover, for any $Z\in\overline{\DM_L}$, the M\"obius transformation $(\Tt^z)^{-1}\cdot Z$ is well-defined and lies in $\overline{\DM_L}$. Furthermore, only for $z\in\DM$,
$$
(\Tt^z)^{-1}\cdot Z\;\in\;\DM_L
\;.
$$
\end{lemma}

\noindent {\bf Proof.}
With the notations of the lemma, the matrix $\Tt=\varphi(S)$ satisfies the defining equation $\Tt^*\Ll\Tt=\Ll$ of the group $\mbox{\rm U}(L,L)$ so that
\begin{equation}
\label{eq-symT}
\begin{pmatrix}
 A & B \\
C & D
\end{pmatrix}^*
\,\Ll \,
\begin{pmatrix}
 A & B \\
C & D
\end{pmatrix} \; = \;
\Ll
\;.
\end{equation}
Thus a direct computation using the identities contained in \eqref{eq-symT} leads to \eqref{eq-Prel}. Note that alternatively $\Pp^z$ can be expressed in terms of $C$ and $D$ only. It remains to show the positivity of $\Pp^z$. Let $\binom{\phi}{\psi}\in\CM^{2L}$ be with, say, $\|\psi\|\geq \|\phi\|$. Then
\begin{eqnarray*}
\binom{\phi}{\psi}^*
\mathcal{P}^z
\binom{\phi}{\psi}
 & = &
\left( |z|^{-2} -1 \right)\,  \phi^*A^*A\phi
+
\left( (\overline{z})^{-1} -z \right)\,  \phi^*A^*B\psi \\
&  & + \;
\left( z^{-1} -\overline{z} \right)\, \psi^*B^*A\phi
+
(1-|z|^2)\, \psi^*B^*B\psi + (1-|z|^2)\,\psi^*\psi
\\
& \geq &
\left( |z|^{-2} -1 \right) \, \phi^*A^*A\phi
\,-\,
2\,\left| (\overline{z})^{-1} -z  \right| \left( \phi^*A^*A\phi \right)^\frac{1}{2} \left(\psi^*B^*B\psi \right)^\frac{1}{2}  
\\
& & + \;
(1-|z|^2) \,\psi^*B^*B\psi + (1-|z|^2)\,\psi^*\psi
\\
& \geq & 
\left( 
\left( |z|^{-2} -1 \right)^\frac{1}{2} \left( \phi^*A^*A\phi \right)^\frac{1}{2}
-
\left(1-|z|^2\right)^\frac{1}{2}\left( \psi^*B^*B\psi \right)^\frac{1}{2}
\right)^2 
+
(1-|z|^2)\;\psi^*\psi \\
& \geq &
\frac{1}{2}\;
(1-|z|^2)\;
\binom{\phi}{\psi}^*\binom{\phi}{\psi}
\;,
\end{eqnarray*}
where in the last inequality the bound  $\|\psi\|\geq \|\phi\|$ was used.
The case $\|\psi\|\leq \|\phi\|$ is dealt in the same manner.

\vspace{.1cm}

As to the last claims, let us begin by noting 
\begin{equation}
\label{eq-transferinverse}
(\Tt^z)^{-1}
\;=\;\Ll\,(\Tt^{\overline{z}^{-1}})^*\,\Ll
\;=\;
\begin{pmatrix}
z\,A^* & -C^*
\\
-B^* & \frac{1}{z}\,D^*
\end{pmatrix}
\;.
\end{equation}
Thus we need to check the invertibility of $-B^*Z+\frac{1}{z}D^*$ for $Z\in\overline{\DM_L}$ and $z\in\DM$. For that purpose let us use again \eqref{eq-symT} so that, in particular, $D^*D-B^*B=\one$. Thus $D^*D\geq 0$ and $D$ is invertible. Also $(BD^{-1})^*BD^{-1}=\one-(D^{-1})^*D^{-1}<\one$ from which follows $\|BD^{-1}\|<1$. Therefore 
$$
-B^*Z+\frac{1}{z}\;D^*\;=\;\frac{1}{z}\;D^*\,\bigl(\one-z\,(BD^{-1})^*Z\bigr)
\;,
$$
is indeed invertible. Thus $(\Tt^z)^{-1}\cdot Z=(zA^*Z -C^*)(-B^*Z+z^{-1}D^*)^{-1}$ is well-defined and
\begin{equation}
 \label{eq-Moebplane}
(\Tt^z)^{-1}
\begin{pmatrix}
Z \\ \one
\end{pmatrix}
\;=\;
\begin{pmatrix}
(\Tt^z)^{-1}\cdot Z \\ \one
\end{pmatrix}
\bigl(-B^*Z+z^{-1}\;D^*\bigr)
\;.
\end{equation}
But using \eqref{eq-Prel}
\begin{eqnarray*}
\begin{pmatrix}
Z \\ \one
\end{pmatrix}^*
\bigl((\Tt^z)^{-1}\bigr)^*\,\Ll\,
(\Tt^z)^{-1}
\begin{pmatrix}
Z \\ \one
\end{pmatrix}
& = &
\begin{pmatrix}
Z \\ \one
\end{pmatrix}^*
\bigl((\Tt^z)^{-1}\bigr)^*\,
\Bigl[
(\Tt^z)^*\,\Ll\,\Tt^z\,-\,\Pp^z
\Bigr]
\,
(\Tt^z)^{-1}
\begin{pmatrix}
Z \\ \one
\end{pmatrix}
\\
& < &
\begin{pmatrix}
Z \\ \one
\end{pmatrix}^*\,\Ll\,
\begin{pmatrix}
Z \\ \one
\end{pmatrix}
\;=\;Z^*Z-\one\;\leq\;0\;,
\end{eqnarray*}
which together with \eqref{eq-Moebplane} shows the last claim.
\hfill $\Box$

\begin{coro}
\label{coro-positive}
With the positive matrix $\mathcal{P}^z_{2k}$ defined in terms of entries of $\mathcal{T}_{2k}^z$ as in {\rm Lemma~\ref{lem-lemmaP}},
$$
\Tt^z(N,0)^*\,\Ll\,\Tt^z(N,0)
\; =\; 
\Ll \;+\;  
\sum_{k=1}^{{N}/{2}} \;(\mathcal{T}^z_1)^* \cdots 
(\mathcal{T}_{2k-1}^z)^*\,\mathcal{P}^z_{2k}\, \mathcal{T}_{2k-1}^z \cdots  \mathcal{T}^z_1 
\;.
$$
\end{coro}

\noindent {\bf Proof.}  
This follows by iterating Lemma~\ref{lem-lemmaP} and using $ (\mathcal{T}^z_{2k-1})^* \Ll\, \mathcal{T}^z_{2k-1} = \Ll$.
\hfill $\Box$

\begin{lemma}
\label{lem-invertible} 
Let $a,b$ be $L \times L$ matrices and set $\Phi=\binom{a}{b}$. If $\Phi^*\Ll\Phi<0$, then $b$ is invertible. On the other hand, if $\Phi^*\Ll\Phi>0$, then $a$ is invertible.
\end{lemma}

\noindent {\bf Proof.}  
 $\Phi^*\Ll\Phi<0$ implies that $a^*a-b^*b <0$. If there is a vector $v$ such that $bv = 0$, then  $v^*a^*av <0$ which is impossible and therefore $b$ is invertible.
If $a^*a-b^*b >0$, the same argument implies invertibility of $a$.
\hfill $\Box$

\vspace{.2cm}

\begin{lemma}
\label{lem-invertible2} 
Let $\Psi$ and $\Phi$ be two $2L\times L$ matrices of maximal rank satisfying $\Psi^*\Ll\Psi=0$ and either $\Phi^*\Ll\Phi>0$ or $\Phi^*\Ll\Phi<0$. Then $\Psi^*\Ll\Phi$ is invertible.
\end{lemma}

\noindent {\bf Proof.}  
The claimed invertibility does not depend on normalization so that we may assume that $\Psi^*\Psi=\one$ and $\Phi^*\Phi=\one$. Now the fact that $\Psi$ is $\Ll$-Lagrangian implies
$$
\Psi\,\Psi^*\;+\;\Ll\,\Psi\,\Psi^*\Ll
\;=\;
\one
\;.
$$
Applying $\Psi^*$ and $\Phi$ to the left and right of this equation shows
$$
\Phi^*\Ll\,\Psi\,\Psi^*\Ll\,\Phi
\;=\;
\one\;-\;\Phi^*\,\Psi\,\Psi^*\,\Phi
\;.
$$
Thus let us argue that $\Phi^*\Psi\Psi^*\Phi<\one$ because this then shows that 
$\Phi^*\Ll\Psi\Psi^*\Ll\Phi$ and thus also $\Psi^*\Ll\Phi$ is invertible. Now $\Phi^*\Psi\Psi^*\Phi$ is given by the squares of the scalar products of the vectors in the planes spanned by $\Phi$ and $\Psi$ and its eigenvalues are thus the squares of the cosines of the principal angles between these planes. There is an eigenvalue $1$ if and only if one angle vanishes and therefore if and only if $\Phi$ and $\Psi$ have a direction in common. This would mean that there are non-vanishing vectors $v,w\in\CM^L$ such that $\Psi v=\Phi w$, which is incompatible with $\Psi^*\Ll\Psi=0$ and $\Phi^*\Ll\Phi>0$ (or $\Phi^*\Ll\Phi<0$).
\hfill $\Box$

\begin{lemma}
\label{lem-solution} 
Suppose $|z|<1$. Given $\xi=(\xi_k)_{k=1,\ldots,N}$ with $\xi_k\in\mbox{\rm Mat}(L,\CM)$, the solution $\phi$ of the equation $(\CMV_N-z) \phi = \xi $ is, for even $n$, given by
\begin{equation}
\label{eq-Greensol}
\binom{\phi_{n}}{\psi_n}
\;=\;
\left(
\mathcal{T}^z_{n} \cdots \mathcal{T}_1^z
\begin{pmatrix} \one \\ \one \end{pmatrix} \phi_1
+ 
\sum_{k=1}^{n/2} 
\mathcal{T}_{n}^z \cdots \mathcal{T}_{2k+1}^z 
\begin{pmatrix}
-z^{-1}\delta_{2k}\beta_{2k}^{-1} & z^{-1}\,\one \\
 -\beta_{2k}^{-1} & 0 
\end{pmatrix}
\begin{pmatrix} \xi_{2k-1} \\ \xi_{2k} \end{pmatrix}
\right)
\;,
\end{equation}
where $\beta_{2k}$ and $\delta_{2k}$ are the entries of $S_{2k}$ and $\phi_1$ has to be chosen in the unique manner such that $\psi_N$ given by {\rm \eqref{eq-Greensol}} satisfies
\begin{equation}
\label{eq-psicond}
\psi_{N}
\;=\;
\begin{pmatrix} V^* \\ 0 \end{pmatrix}^*
\left(
\mathcal{T}^z_{N} \cdots \mathcal{T}_1^z
\begin{pmatrix} \one \\ \one \end{pmatrix} \phi_1
+ 
\sum_{k=1}^{N/2} 
\mathcal{T}_{N}^z \cdots \mathcal{T}_{2k+1}^z 
\begin{pmatrix}
-z^{-1}\delta_{2k}\beta_{2k}^{-1} & z^{-1}\,\one \\
 -\beta_{2k}^{-1} & 0 
\end{pmatrix}
\begin{pmatrix} \xi_{2k-1} \\ \xi_{2k} \end{pmatrix}
\right)
\;.
\end{equation}
%
\end{lemma}

\noindent {\bf Proof.} As above, the equation $(\CMV_N-z) \phi = \xi $ is solved using an auxiliary vector $\psi$  satisfying $\CMVL \psi = z \phi + \xi$ and $ \CMVR\phi = \psi$. This is equivalent that, for $k=1,\ldots,\frac{N}{2}$ and respectively $k=1,\ldots,\frac{N}{2}-1$, 
\begin{equation}
\label{eq-Sinhom}
S_{2k} \begin{pmatrix} \psi_{2k-1} \\ \psi_{2k} \end{pmatrix} 
\;=\;
z \begin{pmatrix} \phi_{2k-1} \\ \phi_{2k} \end{pmatrix} + \begin{pmatrix} \xi_{2k-1} \\ \xi_{2k} \end{pmatrix}
\;,
\qquad
S_{2k+1} \begin{pmatrix} \phi_{2k} \\ \phi_{2k+1} \end{pmatrix} 
\;=\;
\begin{pmatrix} \psi_{2k} \\ \psi_{2k+1} \end{pmatrix}
\;,
\end{equation}
together with the  condition $\psi_1=U\phi_1$ and the constraint that $\psi_N=V\phi_N$ stemming from the other boundary condition. Each of the two equations in \eqref{eq-Sinhom} (the first one divided by $z$) is transformed using Proposition~\ref{prop-Stransferequiv}:
$$
\begin{pmatrix} \phi_{2k} \\ \psi_{2k} \end{pmatrix} 
\;=\;
\Tt^z_{2k}
\begin{pmatrix} \psi_{2k-1} \\ \phi_{2k-1} \end{pmatrix} + 
\begin{pmatrix}
-z^{-1}\delta_{2k}\beta_{2k}^{-1} & z^{-1}\,\one \\
 -\beta_{2k}^{-1} & 0 
\end{pmatrix}
\begin{pmatrix} \xi_{2k-1} \\ \xi_{2k} \end{pmatrix}
\;,
\qquad
\begin{pmatrix} \psi_{2k+1} \\ \phi_{2k+1} \end{pmatrix} 
\;=\;
\Tt^z_{2k+1}\begin{pmatrix} \phi_{2k} \\ \psi_{2k} \end{pmatrix}
\;.
$$
Iterating and replacing $\psi_1=U\phi_1$ gives, for even $n$,
\begin{equation} \label{eq-systemGreen}
\begin{pmatrix}
 \phi_{n}
\\
\psi_{n}
\end{pmatrix}
\;=\;
\mathcal{T}^z_{n} \cdots \mathcal{T}_1^z
\begin{pmatrix} \one \\ \one \end{pmatrix} \phi_1
+ 
\sum_{k=1}^{n/2} 
\mathcal{T}_{n}^z \cdots \mathcal{T}_{2k+1}^z 
\begin{pmatrix}
-z^{-1}\delta_{2k}\beta_{2k}^{-1} & z^{-1}\,\one \\
 -\beta_{2k}^{-1} & 0 
\end{pmatrix}
\begin{pmatrix} \xi_{2k-1} \\ \xi_{2k} \end{pmatrix}
\;,
\end{equation}
while for odd $n$ the entries on the l.h.s. are simply exchanged ({\it cf.} the definition \eqref{eq-Phidef} of $\Phi_n^z$). Now one, moreover, has to satisfy the constraint $V\phi_N=\psi_N$. For that purpose, one takes the last equation for $n=N$ and extracts $\psi_N$ which is then set equal to $V\phi_N$. This leads to equation~\eqref{eq-psicond} which can indeed be solved for $\phi_1$ because the matrix
$$
\begin{pmatrix} \one \\ 0 \end{pmatrix}^*
\mathcal{T}^z_{N} \cdots \mathcal{T}_1^z
\begin{pmatrix} \one \\ \one \end{pmatrix}
$$
is invertible by Lemma~\ref{lem-invertible}. Indeed, Corollary~\ref{coro-positive} shows that $\Phi=\mathcal{T}^z_{N} \cdots \mathcal{T}_1^z \binom{\one}{\one}$ satisfies $\Phi^*\Ll\Phi>0$ which is the hypothesis in Lemma~\ref{lem-invertible}.
\hfill $\Box$

\vspace{.2cm}

\noindent {\bf Proof} of Theorem~\ref{theo-Green}.
By the last claim of Lemma~\ref{lem-lemmaP}, $(\Tt^z_{N-1})^{-1}\cdot\bigl((\Tt^z_N)^{-1}\cdot V^*\bigr)=(\Tt^z_N\Tt^z_{N-1})^{-1}\cdot V^*$ is well-defined and lies in the Siegel disc $\DM_L$. Iterating this shows that
$$
(\Tt^z_2\Tt^z_1)^{-1}\cdot\Bigl(\cdots
\bigl((\Tt^z_N\Tt^z_{N-1})^{-1}\cdot V^*\bigr)\cdots\Bigr)
\;=\;
\Tt^z(N,0)^{-1}\cdot V^*
\;=\;V^*:\Tt^z(N,0)
\;,
$$
exists and lies in $\DM_L$. This shows the first claim. Let us note that the invertibility $V^*C^z_N-A^z_N=V^*(C^z_N-VA^z_N)$ also follows from the identity
$$
C^z_N-VA^z_N
\;=\;
-\;\binom{V^*}{\one}^*\Ll\;\Tt^z_N\cdots\Tt^z_1\binom{\one}{0}
\;,
$$
because $\Phi=\Tt^z_N\cdots\Tt^z_1\binom{\one}{0}$ satisfies $\Phi^*\Ll\Phi>0$ as shows Corollary~\ref{coro-positive}, so that Lemma~\ref{lem-invertible2} applies.

\vspace{.1cm}

The Green matrix $G^z_N(V)$ is  the component $\phi_1$ of the solution of $(\CMV_N(V)-z) \phi = \xi $ with inhomogeneity $\xi_k = \delta_{1,k}\one$. Hence by Lemma~\ref{lem-solution} 
$$
\phi_N
\;=\;
\begin{pmatrix} \one \\ 0 \end{pmatrix}^*
\mathcal{T}^z_{N} \cdots \mathcal{T}_1^z
\left\{
\begin{pmatrix} \one \\ \one \end{pmatrix} 
G^z_N(V)
+ 
(\Tt^z_2\Tt^z_1)^{-1}
\begin{pmatrix}
-z^{-1}\delta_{2}\beta_{2}^{-1} & z^{-1}\,\one \\
 -\beta_{2}^{-1} & 0 
\end{pmatrix}
\begin{pmatrix} \one \\ 0 \end{pmatrix}
\right\}
$$
and
$$
V\phi_N
\;=\;
\begin{pmatrix} 0 \\ \one \end{pmatrix}^*
\mathcal{T}^z_{N} \cdots \mathcal{T}_1^z
\left\{
\begin{pmatrix} \one \\ \one \end{pmatrix} 
G^z_N(V)
+ 
(\Tt^z_2\Tt^z_1)^{-1}
\begin{pmatrix}
-z^{-1}\delta_{2}\beta_{2}^{-1} & z^{-1}\,\one \\
 -\beta_{2}^{-1} & 0 
\end{pmatrix}
\begin{pmatrix} \one \\ 0 \end{pmatrix}
\right\}
\;.
$$
Now as $(\Tt^{z}_n)^{-1}=\Ll(\Tt^{\overline{z}^{-1}}_n)^*\Ll$, one checks that using the identities in $\mbox{U}(2L)_\inv$
$$
(\Tt^z_2\Tt^z_1)^{-1}
\begin{pmatrix}
-z^{-1}\delta_{2}\beta_{2}^{-1} & z^{-1}\,\one \\
 -\beta_{2}^{-1} & 0 
\end{pmatrix}
\begin{pmatrix} \one \\ 0 \end{pmatrix}
\;=\;
\begin{pmatrix}0 \\ z^{-1}\one \end{pmatrix}
\;.
$$
Therefore the above two equations become
%
$$
\phi_N
\;=\;
(A^z_N +B^z_N) \;G^z_N(V)+ z^{-1}B^z_N
\;,
\qquad
V\phi_N
\;=\;
(C^z_N +D^z_N) \;G^z_N(V)+  z^{-1}D^z_N
\;.
$$
%
Now these two equations have to be solved for $G^z_N(V)$. For that purpose, the invertibility of 
$$
(D^z_N-VB^z_N)+(C^z_N-VA^z_N)
\;=\;
-\;\binom{V^*}{\one}^*\Ll\;\Tt^z_N\cdots\Tt^z_1\binom{\one}{\one}
\;,
$$
is needed. It follows from Lemma~\ref{lem-invertible2} because $\binom{V^*}{\one}$ is $\Ll$-Lagrangian and $\Phi=\Tt^z_N\cdots\Tt^z_1\binom{\one}{\one}$ satisfies $\Phi^*\Ll\Phi>0$ as already argued in the proof of Lemma \ref{lem-solution}. Therefore
$$
G^z_N(V)
\;=\;
\frac{1}{z}\;
\Bigl((D^z_N-VB^z_N)+(C^z_N-VA^z_N)\Bigr)^{-1}(VB^z_N-D^z_N)
\;.
$$
Now the formula for $G^z_N(V)$ follows.

\vspace{.1cm}

Finally let us calculate $F^z_N(V)$. We start from
\begin{align}
\imath\,F^z_N(V) \; &
= \; \bigl( \pi_1^* (\CMV_N-z)^{-1} \pi_2 \; \pi_2^*\, S_2\, \pi_1 \; + \; \pi_1 (\CMV_N-z)^{-1} \pi_1 \; \pi_1^* 
\,S_2 \,\pi_1 \bigr) U  \; + \;z\; \pi_1 (\CMV_N-z)^{-1} \;\pi_1  \notag\\
\;  &= \; 
\bigl( \pi_1^* (\CMV_N-z)^{-1} \pi_2 \; \gamma_2 \; + \;  G_N(V) \; \alpha_2 \bigr) \; U \; + \; z \; G^z_N(V). \label{eq-Fformula}
\end{align}
Let us denote $G_{1,2} = \pi_1^* (\CMV_N-z)^{-1} \pi_2$, thus we can calculate $G_{1,2}$ using the same procedure than with $G_N(V)$. Notably, $G_{1,2}$ is the equal to the component $\phi_1$ of the matrix-valued solution of $(\CMV_N-z) \phi = \xi $ with inhomogeneity $\xi_k = \delta_{2,k}\one$. Hence
$$
\phi_N
\;=\;
\begin{pmatrix} \one \\ 0 \end{pmatrix}^*
\mathcal{T}^z_{N} \cdots \mathcal{T}_1^z
\left\{
\begin{pmatrix} \one \\ \one \end{pmatrix} 
G_{1,2}
+ 
(\Tt^z_2\Tt^z_1)^{-1}
\begin{pmatrix}
-z^{-1}\delta_{2}\beta_{2}^{-1} & z^{-1}\,\one \\
 -\beta_{2}^{-1} & 0 
\end{pmatrix}
\begin{pmatrix} 0 \\ \one \end{pmatrix}
\right\}
$$
and
$$
V\phi_N
\;=\;
\begin{pmatrix} 0 \\ \one \end{pmatrix}^*
\mathcal{T}^z_{N} \cdots \mathcal{T}_1^z
\left\{
\begin{pmatrix} \one \\ \one \end{pmatrix} 
G_{1,2}
+ 
(\Tt^z_2\Tt^z_1)^{-1}
\begin{pmatrix}
-z^{-1}\delta_{2}\beta_{2}^{-1} & z^{-1}\,\one \\
 -\beta_{2}^{-1} & 0 
\end{pmatrix}
\begin{pmatrix} 0 \\ \one \end{pmatrix}
\right\}
\;.
$$
Using again $(\Tt^{z}_n)^{-1}=\Ll(\Tt^{\overline{z}^{-1}}_n)^*\Ll$ and the identities in $\mbox{U}(2L)_\inv$, one checks that  
$$
(\Tt^z_2\Tt^z_1)^{-1}
\begin{pmatrix}
-z^{-1}\delta_{2}\beta_{2}^{-1} & z^{-1}\,\one \\
 -\beta_{2}^{-1} & 0 
\end{pmatrix}
\begin{pmatrix} 0 \\ \one \end{pmatrix}
\;=\;
\begin{pmatrix}-U^*\gamma_2^{-1} \\ -z^{-1}\alpha_2\gamma_2^{-1} \end{pmatrix}.
$$
Therefore the above two equations become
$$
\phi_N
\;=\;
(A^z_N \;+\;B^z_N)\;G_{1,2} \;-\;A^z_N \,U^*\gamma_2^{-1} \;-\;   z^{-1}\,B^z_N \,\alpha_2\,\gamma_2^{-1}
\;,
$$
and
$$
V\phi_N
\;=\;
(C^z_N \;+\;D^z_N)\;G_{1,2}  \;-\;C^z_N\, U^*\, \gamma_2^{-1} \;-\;   z^{-1}\,D^z_N\, \alpha_2\,\gamma_2^{-1}
\;.
$$
Solving yields 
$$
G_{1,2} \; = \; \bigl[(D^z_N-VB^z_N)+(C^z_N-VA^z_N)\bigr]^{-1}
\bigl[ (D^z_N-VB^z_N)(-z^{-1}\alpha_2\gamma_2^{-1}) - (C^z_N-VA^z_N)U^*\gamma_2^{-1} \bigr]
\;.
$$
Replacing this and $G_N(V)$  in the form \eqref{eq-Green1} into  \eqref{eq-Fformula} leads to the desired formula for $F^z_N(V)$.
\hfill $\Box$

\section{Weyl theory}

For $z\in\DM$, let us set 
$$\mathcal{Q}^z_N 
\;=\; 
(\mathcal{T}^z_N \cdots \mathcal{T}_1^z)^*\,\mathcal{L}\, 
\mathcal{T}^z_N \cdots \mathcal{T}_1^z
\;.
$$
We consider $\Qq^z_N$ as quadratic form on $\CM^{2L}$ and are particularly interested in its maximally isotropic subspaces, also called Lagrangian subspaces. It will shortly be shown that the signature of $\Qq^z_N$ is $(L,L)$ so that the dimension of these Lagrangian subspaces is $L$. By definition the Weyl surface is then the image of (the inverse Cayley transform of) these subspaces under an adequate chart on the Grassmannian called the stereographic projection. Recall that the Grassmannian $\mathbb{G}_L$ of all $L$-dimensional subspaces of $\CM^{2L}$ can be introduced as set of equivalent classes of  $2L \times L$ matrices of maximal rank with respect to the equivalence relation $\Phi \sim \Psi \Longleftrightarrow \Phi = \Psi c$ for some $c\in\mbox{\rm Gl}(L,\CM)$. Elements of $\mathbb{G}_L$ will be denoted by $[\Phi]_\sim$. Let us consider the subset  $\mathbb{G}^{\mbox{\tiny \rm inv}}_L \subset \mathbb{G}_L$ of subspaces represented by some $\Phi = \binom{a}{b} $ with an invertible $L\times L$ matrix $b$. This set is the domain of the stereographic projection $\pi : \mathbb{G}^{\mbox{\tiny\rm inv}}_L \to \mbox{\rm Mat}(L,\CM) $ defined by
\begin{equation}
\label{eq-defpi}
\pi ([\Phi]_\sim) \;=\; a\,b^{-1}\;, 
\qquad \Phi \;=\; \binom{a}{b}
\;.
\end{equation}

\begin{defini}
The Weyl surface is defined by
$$
\partial \Weyl_N^z 
\; =\;  \pi \bigl(\left\{ [\Cc^*\Phi]_\sim \in \mathbb{G}_L \;\big|\; \Phi\; \mbox{\rm isotropic for } \mathcal{Q}_N^z \right\}\bigr)
\;,
$$
and the closed Weyl disc by
$$
\Weyl_N^z 
\; =\;  \pi \bigl(\left\{ [\Cc^*\Phi]_\sim \in \mathbb{G}_L \;\big|\; \Phi^*\mathcal{Q}_N^z\Phi\leq 0 
\right\}\bigr)
\;.
$$

\end{defini}

Of course, we have to check below that this is well-defined, namely that $[\Cc^*\Phi]_\sim$ is in the domain of $\pi$ for all Lagrangian subspaces of $\mathcal{Q}_N^z$. Then the next aim will be to show that the Weyl surface is the surface of a matrix ball and this will ultimately allow to prove estimates on the dependence of $G^z_N(V)$ and $F^z_N(V)$ on the boundary condition $V\in\mbox{\rm U}(L)$.

\begin{proposi} 
\label{prop-Qformulas}
The quadratic form $\mathcal{Q}^z_k$ has the following properties: 

\vspace{0.1cm}

\noindent {\rm (i)} $\mathcal{Q}^z_{2k+1}= \mathcal{Q}^z_{2k}$ and $\mathcal{Q}^z_{2k}  =  
\mathcal{Q}^z_{2k-1} + (\mathcal{T}^z_1)^* \cdots (\mathcal{T}_{2k-1}^z)^*\mathcal{P}^z_{2k} \mathcal{T}_{2k-1}^z \cdots  \mathcal{T}^z_1$. Furthermore 
$$
\mathcal{Q}^z_{N}\; =\; \Ll \,+  \,
\sum_{k=1}^{N/2} (\mathcal{T}^z_1)^* \cdots 
(\mathcal{T}_{2k-1}^z)^*\,\mathcal{P}^z_{2k}\, \mathcal{T}_{2k-1}^z \cdots  \mathcal{T}^z_1 
$$

where $\mathcal{P}^z_{2k}$ is positive matrix given in term of entries of $\mathcal{T}_{2k}^z$ defined in {\rm Lemma \ref{lem-lemmaP}}.

\vspace{0.1cm}

\noindent {\rm (ii)}
$(Q^z_N)^{-1}=\Ll\,Q^{\overline{z}^{-1}}_N\,\Ll$

\vspace{0.1cm}

\noindent {\rm (iii)} $\mathcal{Q}^z_{2k} >  \mathcal{Q}^z_{2k-1}$

\vspace{0.1cm}

\noindent {\rm (iv)}
${\rm signature}(Q^z_N)=(L,L)$
\end{proposi}

\noindent {\bf Proof.}
(i) is just a reformulation of Lemma \ref{lem-lemmaP} and Corollary~\ref{coro-positive}. (ii) results from \eqref{eq-transferinverse}. (iii) is a consequence of (i) and $\mathcal{P}^z_{2k} >0$. Finally, from the definition of $\mathcal{Q}^z_N$, one deduces that 
$$
\mbox{signature}(Q^z_N)\;=\; \mbox{signature}(\Ll)\;=\; (L,L)\;,
$$
which is (iv).
\hfill $\Box$


\begin{proposi}
\label{prop-isovector}
Let $E^z_N(V)$ be given as in {\rm Theorem~\ref{theo-Green}}. 
%
%
If $\Phi$ is Lagrangian for $\mathcal{Q}^z_N$, then there is a unique unitary $V$ and an invertible matrix $b$ with
$$
\Phi 
\;=\; 
\begin{pmatrix}
E_N^z(V) \\ \one 
\end{pmatrix}
\;b\;.
$$
Moreover, $[\Cc^*\Phi]_\sim$ is in the domain of $\pi$.
\end{proposi}

\noindent {\bf Proof.} 
Let $\Phi=\binom{a}{b}$ satisfy $\Phi^*\mathcal{Q}^z_N \Phi=0$. This implies that $\Phi^*\Ll \Phi < 0$ by Proposition~\ref{prop-Qformulas}(i) and thus by Lemma~\ref{lem-invertible} that $b$ is invertible. As $\binom{a}{b}$ is isotropic if and only if $\binom{ab^{-1}}{\one}b$ is isotropic, we may assume that $b=\one$. Now $\Phi^*\mathcal{Q}^z_N \Phi=0$ is equivalent to
$$
(A^z_Na + B^z_N  )^*(A^z_Na + B^z_N  ) 
\;=\; 
(C^z_Na + D^z_N  )^*(C^z_Na + D^z_N  ) \;.
$$
Using polar decomposition, it thus follows that there exists a unique unitary $V$ such that 
$$
(VA^z_N - C^z_N )a \;=\; 
D^z_N - VB^z_N
\;.
$$
As $VA^z_N - C^z_N$ is invertible by Theorem~\ref{theo-Green}, it follows indeed that $a=E^z_N(V)$. Furthermore,
\begin{equation}
\label{eq-beforepi}
\Cc^*\Phi
\;=\;
\begin{pmatrix}
E_N^z(V)+\one \\
\imath(E_N^z(V)-\one)
\end{pmatrix}
\,b\;.
\end{equation}
By Theorem~\ref{theo-Green}, $E_N^z(V)$ is in the Siegel disc so that $E_N^z(V)-\one$ is invertible and therefore $[\Cc^*\Phi]_\sim$ is in the domain of $\pi$.
\hfill $\Box$


\begin{theo} 
\label{theo-Weyl1}
One has
$$
\partial \Weyl_N^z \; 
\; = \;  \left\{  F^z_N(V) \,\left|  
\, V \in \mbox{\rm U}(L)\right. \right\}
\;  = \; 
\left\{ \imath(\one+2\,z\, G^z_N(V))
\,\left|  
\, V \in \mbox{\rm U}(L)\right. \right\}
\;.
$$
\end{theo}

\noindent {\bf Proof.} The first equality follows immediately from \eqref{eq-beforepi} and \eqref{eq-resol1} in Theorem~\ref{theo-Green}. The second one then follows by combining \eqref{eq-resol1} and \eqref{eq-Green1}. 
\hfill $\Box$

\vspace{.2cm}

The next aim is to analyze the geometry of the Weyl surface. This can be done in complete analogy with \cite{SB3} if one works with the Cayley transform of the quadratic form:
\begin{equation}
\label{eq-Qtilde}
\WQ^z_N
\;=\;
\Cc^*\Qq^z_N\,\Cc
\;=\;
\imath\,\Jj\;+\;
\sum_{k=1}^{N/2} 
\;\Cc^*\,\Tt^z(2k-1,0)^*\,\Pp^z_{2k} \,\Tt^z(2k-1,0)\,\Cc
\;,
\end{equation}
where
$$
\Jj
\;=\;
\frac{1}{\imath}\;
\Cc^*\Ll\,\Cc
\;=\;
\begin{pmatrix}
0 & -\one \\ \one & 0
\end{pmatrix}
\;.
$$

\begin{defini}
\label{def-radial}
The radial and central operators are defined by
$$
R_{N}^z \; = \; \left[\binom{\one}{0}^* \WQ^z_N \binom{\one}{0} \right]^{-1}\;,
\qquad
S_{N}^z \;  = \; -\, R_{N}^z\;\binom{\one}{0}^* \WQ^z_N \binom{0}{\one}\;.
$$
\end{defini}

\begin{proposi}
\label{prop-SNRN} Let $z\in\DM$. Then $R_{N}^z$ is well-defined, positive and decreasing in $N$. Also $R^{\overline{z}^{-1}}_N$ is well-defined, but negative. Moreover:
$$ 
 (S_{N}^{z})^*\; = \;  S_{N}^{\overline{z}^{-1}}
\;,
\qquad
\binom{0}{\one}^* \WQ^z_N \binom{0}{\one}  
\;=\; 
(S_{N}^z)^* (R^z_{N})^{-1} S^z_{N} \;+\;    R_{N}^{\overline{z}^{-1}} 
\;.
$$
\end{proposi}

\noindent {\bf Proof.} All claims on $R^z_N$ follow by taking the matrix element of \eqref{eq-Qtilde} because $\binom{\one}{0}^*\Jj\binom{\one}{0}=0$ and the map $r\in(0,\infty)\mapsto -r^{-1}$ is operator monotone. For the second claim, one has to adapt Lemma~\ref{lem-lemmaP} and Corollary~\ref{coro-positive} to the case $|z|>1$, but then the proof is identical. Next let us note that the Cayley transform of Proposition~\ref{prop-Qformulas}(ii) reads
$$
\Jj\;=\;
\WQ^{\overline{z}^{-1}}_N \Jj \WQ^z_N 
\;.
$$
%
Using $\mathcal{J} =  \binom{0}{\one} \binom{\one}{0}^* -  \binom{\one}{0}\binom{0}{\one}^*$, this yields 
$$
\Jj\; =  \; 
\WQ^{\overline{z}^{-1}}_N \binom{0}{\one} \binom{\one}{0}^* \WQ^z_N
\;-\;
\WQ^{\overline{z}^{-1}}_N\binom{\one}{0}\binom{0}{\one}^* \WQ^z_N
\;.
$$
The upper left and upper right entries of this equation give
\begin{align*}
0 & \;=  \;
-\,(R_N^{{\overline{z}}^{-1}})^{-1} S_N^{{\overline{z}}^{-1}} (R_N^{z})^{-1} 
\;+\; (R_N^{{\overline{z}}^{-1}})^{-1} (S_N^{z})^* (R_N^{z})^{-1}
\;,
\\
-\,\one & \;=\;
(R_N^{{\overline{z}}^{-1}})^{-1} S_N^{{\overline{z}}^{-1}} (R_N^{z})^{-1} S^z_N
\;-\;
(R_N^{{\overline{z}}^{-1}})^{-1} 
\binom{0}{\one}^*
\WQ^{z}_N
\binom{0}{\one}
\;.
\end{align*}
These two equations lead to the remaining two claims.
\hfill $\Box$

\begin{theo}
\label{theo-Weylsurf}
The Weyl discs are strictly nested matrix discs, namely one has $\Weyl^z_{N}\subset \Weyl^z_{N-1}$ and $\partial \Weyl^z_{N-1}\cap \Weyl^z_{N+1}=\emptyset$, as well as
$$
\partial \Weyl_N^z  
\;=\; \left\{\left. S_{N}^z +  \; (R_{N}^z )^{\frac{1}{2}} W (-R_{N}^{\overline{z}^{-1}})^{\frac{1}{2}}\,\right|\, 
W \in \mbox{\rm U}(L) \right\}\;.
$$
\end{theo}

\noindent {\bf Proof.} 
The first claim is an immediate corollary of Proposition~\ref{prop-Qformulas}(iii). Let $\Phi_E = \binom{E}{1}$ be a $\Qq_N^z-$Lagrangian plane. This is equivalent to
\begin{eqnarray*}
0 & = &
 \binom{E+\one}{\imath(E-\one)}^* \WQ_N^z  \binom{E+\one}{\imath(E-\one)} 
\\
& = & 
(E+\one)^*(R^z_N)^{-1}(E+\one)
\;+\;\imath\,(E-\one)^*(S^z_N)^*(R^z_N)^{-1}(E+\one)
\\
& &
\;-\;\imath\,(E+\one)^*(R^z_N)^{-1} S^z_N(E-\one)
\;+\;
(E-\one)^*\bigl(
(S_{N}^z)^* (R^z_{N})^{-1} S^z_{N} \;+\;    R_{N}^{\overline{z}^{-1}}
\bigr)
(E-\one)
\;,
\end{eqnarray*}
where in the second equality Definition~\ref{def-radial} and Proposition~\ref{prop-SNRN} were used. Rewriting gives
$$
\Bigl((E+\one) - \imath \,S_N^z (E-\one) \Bigr)^* (R_N^z)^{-1}  \Bigl((E+\one) + \imath\,S_N^z (E-\one) \Bigr) 
\; = \;
 - \,(E-\one)^* R_N^{\overline{z}^{-1}} (E-\one) 
$$
By Proposition~\ref{prop-isovector} and Theorem~\ref{theo-Green}, $E\in\DM_L$ so that $E-\one$ is invertible. Hence
$$
\Bigl(-\,\imath(E+\one)(E-\one)^{-1} - S_N^z  \Bigr)^* (R_N^z)^{-1}  \Bigl(-\,\imath(E+\one)(E-\one)^{-1} - S_N^z  \Bigr)  \; = \; - R_N^{\overline{z}^{-1}}  
\;.
$$
Therefore there exists a unique unitary $W$ such that 
$$
\frac{1}{\imath}\;(E+\one)(E-\one)^{-1}\; =\; S_N^z +  \; (R_N^z )^{\frac{1}{2}} W 
(-  R_N^{\overline{z}^{-1}})^{\frac{1}{2}} \;.
$$
As the l.h.s. is precisely $\pi([\Cc^* \Phi_E]_\sim)$, this concludes the proof.
\hfill $\Box$

\vspace{.2cm}

The following is now an immediate consequence.

\begin{coro}
\label{coro-Gweyl}
For any boundary condition $V$ and $z\in\DM$, there is a unitary $W\in\mbox{\rm U}(L)$ such that
$$
F_N (V) \; = \;   S_{N}^z \; +\;( R_{N}^z )^{\frac{1}{2}} W  (- R_{N}^{\overline{z}^{-1}})^{\frac{1}{2}}
\;, 
$$
and
$$
G_N^z (V) \; = \; \frac{1}{2z} \; \Bigl(\one\,-\, \imath\,S_{N}^z \,- \;\imath\, (R_{N}^z )^{\frac{1}{2}} W  (- R_{N}^{\overline{z}^{-1}})^{\frac{1}{2}}\Bigr)
\;. 
$$
\end{coro}

By now the beautiful theory of Weyl discs is complete. Let us finally come to its main analytical application, namely the control of the dependence of the Green matrix $G_N^z (V)$ and the resolvent $F_N^z (V)$ (both matrix elements of the resolvent at the boundary site $1$) on the boundary condition $V$ on the (other) boundary site $N$. Let us start by noting the following.

\begin{coro}
\label{coro-GFbound}
For any $V, V'\in\mbox{\rm U}(L)$,
$$
\| F_N^z (V) -F_N^z (V')\|^2 \;\leq \; \|R_{N}^z \| \,\| R_{N}^{\overline{z}^{-1}} \|
\;,
\qquad
\| G_N^z (V) -G_N^z (V')\|^2\; \leq\;  \frac{1}{4\,|z|^2} \;\|R_{N}^z \| \,\| R_{N}^{\overline{z}^{-1}} \|
\;.
$$
\end{coro}

This becomes useful in combination with the following result.

\begin{proposi}
\label{prop-Rbound}
One has
$$
\max\left\{
\|R_{N}^z \| \, , \, \|R_{N}^{\overline{z}^{-1}} \|
\right\}
\; \leq \;  
\frac{1}{(1 - |z|^2)^{2}}\;\frac{8}{N} 
\;.
$$
\end{proposi}

\noindent {\bf Proof.} Using \eqref{eq-Qtilde} and the bound from Lemma~\ref{lem-lemmaP},
\begin{eqnarray*}
\binom{\one}{0}^* \WQ^z_N \binom{\one}{0} 
& = & 
\frac{1}{2}\;\sum_{k=1}^{N/2} \binom{\one}{\one}^* (\mathcal{T}^z_{1})^*\cdots (\mathcal{T}^z_{2k-1})^*\,\mathcal{P}^z_{2k} \,\mathcal{T}^z_{2k-1}\cdots \mathcal{T}^z_{1} \binom{\one}{\one} 
\\
& \geq & 
\frac{1-|z|^2}{4}\; 
\sum_{k=1}^{N/2} (\Phi^z_{2k-1})^* \Phi^z_{2k-1} 
\; .
\end{eqnarray*}
Now $(\Phi^z_1)^*\Phi^z_1=2\,\one$ and for $k\geq 2$
$$
(\Phi^z_k)^* \Phi^z_k \; \geq \;  (\Phi^z_k)^* \,\Ll\, \Phi^z_k 
\; = \; \binom{\one}{\one}^* \Qq^z_k \binom{\one}{\one} 
\; \geq \; 
\binom{\one}{\one}^* (\Tt^z_1)^*\,\Pp_2^z\,\Tt^z_1\, \binom{\one}{\one}
\;\geq\;
(1-|z|^2)\;\one
\;.
$$
(This is rough, but sufficient for our purposes.) Therefore
$$
\binom{\one}{0}^* \WQ^z_N \binom{\one}{0} 
\; \geq \; 
\frac{N}{8} \; \bigl(1-|z|^2\bigr)^2\;\one
\;,
$$
from which the upper bound on $R_{N}^{z}$ follows. Using a version of Lemma~\ref{lem-lemmaP} for $z\not\in\overline{\DM}$, the bound on $R_{N}^{\overline{z}^{-1}}$ is shown similarly.
\hfill $\Box$

\section{Spectral measures of semi-infinite scattering zippers}

Let now $\CMV$ be a semi-infinite scattering zipper operator associated to a sequence $(S_n)_{n\geq 2}$ of scattering matrices $S_n\in\mbox{U}(2L)_\inv$ and a boundary condition $U\in \mbox{\rm U}(L)$. Then it follows from Proposition~\ref{prop-Rbound} combined with Corollary~\ref{coro-GFbound} that the limit
$$
F^z\;=\;
\lim_{N\to\infty}\;F^z_N(V)
\;,
$$
exists and is independent of the choice of $V$. In the terminology of Weyl theory, a semi-infinite scattering zipper operator is always in the limit point case. Moreover, the convergence is uniform in $z$ on compact subsets of $\DM$. Therefore, $z\in\DM\mapsto F^z$ is analytic. As $\Im m(F^z_N(V))>0$ and $F^0_N(V)=\imath\,\one$ it follows that also $\Im m(F^z)>0$ and $F^0=\imath\,\one$. Therefore the Riesz-Herglotz representation theorem recalled in Appendix B can be applied to show that there is a unique matrix-valued probability measure $\mu$ on $\SM^1$ such that
$$
F^z\;=\;
\imath\,\int_{\SM^1} \mu(d\xi)\;\frac{\xi+z}{\xi-z}
\;.
$$
This measure is called the spectral measure of $\CMV$. It dominates all other spectral measure (because the range of $\pi_1$ is a cyclic subspace for $\CMV$). Resuming, there is a map $\vartheta$ that associates to every semi-infinite scattering zipper operator a matrix-valued probability measure $\mu=\vartheta(\CMV)$ on $\SM^1$. There is still some gauge freedom though allowing several operators $\CMV$ to have the same spectral measure, as already pointed out in \cite{BB,BHJ}. This is dealt with in the following result.

\begin{theo}
\label{theo-bijection}
Let $\Mm(U)$ denote the set of all semi-infinite scattering zipper operators with left boundary condition $U\in\mbox{\rm U}(L)$ and scattering matrices $(S_n)_{n\geq 2}$ given by $S_n=S(\alpha_n,U_n,V_n)$ where $\|\alpha_n\|<1$ and $U_n,V_n\in\mbox{\rm SU}(L)$. Then the map $\vartheta$ establishes a bijection between $\Mm(U)$ and the matrix-valued probability measures on $\SM^1$.
\end{theo}

For the proof, it remains to construct an inverse to $\vartheta$, namely to construct a scattering zipper with boundary condition $U$ from a given matrix-valued probability measure $\mu$ on $\SM^1$, such that $\mu$ is again its spectral measure. Adapting the approach in \cite{DPS}, this will be done by a Gram-Schmidt procedure associated an adequate scalar product $\langle  \,.\,,\,.\, \rangle$ (with values in the $L\times L$ matrices) and an adequate basis for the functions on $\SM^1$ (this will be given by Laurent polynomials in an adequate order as suggested by \cite{CMV}). Let us define 
\begin{equation}
\label{eq-prodscal}
\langle  f,g \rangle \; = \; \int_{\SM^1} g(z) \; \mu(dz) \;  f^*(z)
\;,
\end{equation}
where $f$ and $g$ are matrix-valued functions. At first sight, there seems to be something wrong with this definition. Indeed, one can define another scalar product by exchanging $f$ and $g$ on the r.h.s.. When the Gram-Schmidt procedure below is done w.r.t. this other scalar product, one obtains different orthonormal polynomials, and only their so-called Szeg\H{o} transformation \cite{DPS} will lead to the polynomials $\phi^z$ and $\psi^z$ used in Section~\ref{sec-transfer}. Somewhat anticipating, already the notations $\phi^z$ and $\psi^z$ will be chosen already now.  Using \eqref{eq-prodscal} allows to avoid using the Szeg\H{o} transformation and also simplifies the calculations below. The product \eqref{eq-prodscal} is left matrix-linear in the second argument and anti-linear in the first, namely it satisfies for matrix valued functions $f$, $g$ and $h$ on $\SM^1$ as well as a matrix $\alpha$ that
$$
\langle  f, g+\alpha \,h \rangle\;=\;\langle  f,g \rangle+\alpha \;\langle  f,h \rangle \;, 
\qquad  
\langle f+ \alpha\, g,h \rangle\;=\;\langle  f,h \rangle+\langle  g,h \rangle \;\alpha^* 
\;.
$$
Now two sequences $\phi^z=(\phi^z_n)_{n\geq 1}$ and $\psi^z=(\psi^z_n)_{n\geq 1}$ of orthonormal families of matrix-valued Laurent polynomials in $z$ will be constructed by the Gram-Schmidt algorithm w.r.t. \eqref{eq-prodscal}. For $\phi^z$, one orthonormalizes the sequence $\{\one,z^{-1}\one,z^1\one,z^{-2}\one,z^{2}\one, \dots \}$. Then $\phi^z_1=\one$ and the $n$th element of the resulting orthonomal sequence is $\phi^z_n$. Simarly, $\psi^z$ is obtained by orthonormalizing $\{U,z\one,z^{-1}\one,z^2\one,z^{-2}\one, \dots \}$. Hence $\psi^z_1=U$. The orthonormality relations read: 
$$
\langle \phi^z_m, \phi^z_n \rangle \; =\; \delta_{m,n}\,\one\;, 
\qquad 
\langle \psi^z_m, \psi^z_n \rangle \; =\; \delta_{m,n}\,\one\;.
$$
Let us note that the $\phi^z$ satisfy
\begin{equation}
\label{eq-lead1}
\phi^z_{2n} \;=\;  {\kappa}_{2n}\, z^{-n} + {p}(z)\;,
\qquad
\phi^z_{2n+1} \;=\; {\kappa}_{2n+1} \,z^{n} + {q}(z)\;,
\end{equation}
where the leading coefficient ${\kappa}_{n}$ are invertible matrices, ${p}$ is a matrix polynomial in the span of $z^{n-1},\dots,z^{-n+1}$ and ${q}$ a polynomial in the span of $z^{n-1},\dots ,z^{-n}$. Similarly,
\begin{equation}
\label{eq-lead2}
\psi^z_{2n} \;=\; \widetilde{\kappa}_{2n}\, z^n + \widetilde{p}(z)\;,
\qquad
\psi^z_{2n+1} \;=\; \widetilde{\kappa}_{2n+1} z^{-n}+\widetilde{q}(z)\;,
\end{equation}
with invertible $\widetilde{\kappa}_n$ and polynomials $\widetilde{p}$ and $\widetilde{q}$ in the span of $z^{n-1}, \dots ,   z^{-n+1}$ and $z^{n},\dots,z^{-n+1}$ respectively. Now let us also define
$$
\rho_n \; =  \; \widetilde{\kappa}_{n-1} ({\kappa}_n)^{-1}\;, 
\qquad \widetilde{\rho}_n \; = \; {\kappa}_{n-1} (\widetilde{\kappa}_n)^{-1}\;.
$$
%

\begin{lemma} 
\label{lem-zsegorecurs}
The orthonormal polynomials $\phi^z$ and $\psi^z$ have the following properties.

\vspace{.1cm} 

\noindent {\rm (i)}
There exist $L \times L$  matrices $\alpha^1_{2n+1}, \alpha^2_{2n+1}, \alpha^3_{2n+2}, \alpha^4_{2n+2}$, such that
\begin{equation}\label{eq-zsego1}
\psi^z_{2n} - \rho_{2n+1} \phi^z_{2n+1} \; =\; \alpha^1_{2n+1} \phi^z_{2n}\;,
\qquad
\phi^z_{2n} - \widetilde{\rho}_{2n+1} \psi^z_{2n+1} \; =\; \alpha^2_{2n+1} \psi^z_{2n}\;,  
\end{equation}

and
\begin{equation}
\label{eq-zsego3}
z^{-1} \psi^z_{2n-1} - \rho_{2n}\, \phi^z_{2n} \; =\; \alpha^3_{2n}\, \phi^z_{2n-1}\;,  
\qquad
z \phi^z_{2n-1} - \widetilde{\rho}_{2n} \,\psi^z_{2n} \; =\; \alpha^4_{2n}\, \psi^z_{2n-1}\;.  
\end{equation}

\noindent {\rm (ii)} One has
$$
\alpha^1_{2n+1} \; = \; (\alpha^2_{2n+1} )^*\;,
\qquad  
\alpha^3_{2n} \; = \; (\alpha^4_{2n})^*\;. 
$$

In the following, we will simply denote $\alpha^1_{2n+1}$ by  $\alpha_{2n+1}$ and $\alpha^3_{2n}$ by $\alpha_{2n}$.

\vspace{.1cm} 

\noindent {\rm (iii)} There are unique unitaries $U_n$ and $ V_n$ with unit determinant such that
\begin{equation}
\label{eq-rhoUV}
 \rho_{n} \;=\; (\one-\alpha_{n}\alpha_{n}^*)^{\frac{1}{2}} U_{n}\;
,
\qquad
 \widetilde{\rho}_{n} \;=\; (\one-\alpha_{n}^*\alpha_{n})^{\frac{1}{2}} V_{n}^*\;.
\end{equation}

\vspace{.1cm} 

\noindent {\rm (iv)} One has $\alpha_n^* \alpha_n<\one$.
\end{lemma}

\noindent {\bf Proof.} All four relations in \eqref{eq-zsego1} and \eqref{eq-zsego3} are dealt in the same manner, so let us focus on the first one. By \eqref{eq-lead1} and \eqref{eq-lead2}, $\psi^z_{2n} - \rho_{2n+1} \phi^z_{2n+1}$ is in the span of  $z^{n-1}, \dots ,   z^{-n}$. Moreover, this polynomial  is orthonormal to $z^{n-1}\one, \dots ,   z^{-n+1} \one$ as $\psi^z_{2n}$ and $\phi^z_{2n+1}$ are by construction. This implies that $\psi^z_{2n} - \rho_{2n+1} \phi^z_{2n+1}$ is a left multiple of $\phi^z_{2n}$. 

\vspace{.1cm}

For item (ii), let us first calculate as follows: 
$$
\alpha^1_{2n+1} \; = \; \alpha^1_{2n+1} \langle \phi^z_{2n}, \phi^z_{2n} \rangle 
\; = \; \langle \phi^z_{2n},\alpha^1_{2n+1}\phi^z_{2n} \rangle 
\; = \; \langle \phi^z_{2n}, \psi^z_{2n} - \rho_{2n+1} \phi^z_{2n+1}  \rangle 
\;=\;  \langle \phi^z_{2n}, \psi^z_{2n} \rangle\;.
$$
On the other hand,
$$
\alpha^2_{2n+1} \; = \; \alpha^2_{2n+1} \langle \psi^z_{2n}, \psi^z_{2n} \rangle 
\;= \; \langle \psi^z_{2n},\alpha^2_{2n+1}\psi^z_{2n} \rangle
\;= \; \langle \psi^z_{2n}, \phi^z_{2n} - \widetilde{\rho}_{2n+1} \psi^z_{2n+1}  \rangle 
\;= \; \langle \psi^z_{2n}, \phi^z_{2n} \rangle\;.
$$
Thus $\alpha^2_{2n+1} = (\alpha^1_{2n+1})^*$. The other equality is checked in a similar manner. 

\vspace{.1cm}

Next let us check the first identity of  \eqref{eq-rhoUV} for odd index:
\begin{align*}
\one \; &= \; \langle \psi^z_{2n}, \psi^z_{2n} \rangle \\
        &= \; \langle \rho_{2n+1} \phi^z_{2n+1} + \alpha_{2n+1} \phi^z_{2n}, \rho_{2n+1} \phi^z_{2n+1} + \alpha_{2n+1} \phi^z_{2n} \rangle \\
        &= \;  \langle \rho_{2n+1} \phi^z_{2n+1} , \rho_{2n+1} \phi^z_{2n+1}   \rangle \;+\;  \langle \alpha_{2n+1} \phi^z_{2n}, \alpha_{2n+1} \phi^z_{2n} \rangle \\
        &= \; \rho_{2n+1} \langle  \phi^z_{2n+1} ,  \phi^z_{2n+1}    \rangle \rho_{2n+1}^* \;+\; \alpha_{2n+1} \langle  \phi^z_{2n},  \phi^z_{2n} \rangle \alpha_{2n+1}^* \\
        &= \;  \rho_{2n+1}  \rho_{2n+1}^* + \alpha_{2n+1}\alpha_{2n+1}^*
        \;.
\end{align*}
Thus, there exists a unique unitary of unit determinant denoted $U_{2n+1}$ such that 
$$
 \rho_{2n+1}\; =\; (\one-\alpha_{2n+1}\alpha_{2n+1}^*)^{\frac{1}{2}}\, U_{2n+1}\;.
$$
The other cases are dealt with in the same way. As each $\rho_n$ and $\widetilde{\rho}_n$ is invertible, the identities \eqref{eq-rhoUV} also imply (iv).
\hfill $\Box$

\vspace{.2cm}

\noindent {\bf Proof} of Theorem \ref{theo-bijection}. From the density of Laurent polynomials, it follows that  $(\phi^z_n)_{n\geq 1}$ and  $(\psi^z_n)_{n \geq 1}$ are orthonormal basis of the square integrable matrix-valued functions on $\SM^1$ w.r.t. $\langle\,.\,|\,.\,\rangle$. Thus any  $L \times L$ matrix valued function $f$ can be expanded as follows: 
\begin{equation}
\label{eq-expand}
f \;=\; \sum_{n=1}^\infty  \,\langle  f,\phi^z_n \rangle \,\phi^z_n
\;=\; \sum_{n=1}^\infty  \,\langle f, \psi^z_n \rangle \,\psi^z_n\;.
\end{equation}

Now let us define the matrix entries of semi-infinite matrix-valued operators by
$$
\CMV_{n,m} \;=\; \langle \phi^z_m , z\phi^z_n \rangle
\;,
\qquad
\CMVL_{n,m} \;=\; \langle z\phi^z_n , \psi^z_m \rangle
\;,
\qquad
\CMVR_{n,m} \;=\; \langle \phi^z_m , \psi^z_n \rangle
\;.
$$
Then it follows from \eqref{eq-expand} that
$$
\CMV_{n,m}  
\;=\;
  \langle \phi^z_m, \; \sum_{k=1}^\infty \langle z\phi^z_n , \psi^z_k\rangle \psi^z_k \rangle   \;=\; \sum_{k=1}^\infty \langle z\phi^z_n , \psi^z_k\rangle \langle\phi^z_m, \psi^z_k \rangle \;= \;
 \sum_{k=1}^\infty \CMVL_{n,k}\; \CMVR_{k,m}\;.
$$
It remains to show that $\CMVL$ and $\CMVR$ defined above have the same structure as their homonyms from Section~\ref{sec-transfer}. For $n \geq 1$ it follows from \eqref{eq-zsego1} that
$$
\psi^z_{2n}\; = \;\alpha_{2n+1} \phi^z_{2n} + \rho_{2n+1} \phi^z_{2n+1}\;,
\qquad 
\psi^z_{2n+1}\; =\; (\widetilde{\rho}_{2n+1})^* \phi^z_{2n} - V_{2n+1}\alpha_{2n+1}^* U_{2n+1} \phi^z_{2n+1} 
\;,
$$
where the identity $(\widetilde{\rho}_{2n+1})^{-1}\alpha_{2n+1} \rho_{2n+1} =   V_{2n+1}\alpha_{2n+1}^* U_{2n+1}$ was used. Recalling the notation \eqref{eq-Smatdef}, it follows that
$$
\binom{\psi^z_{2n}}{\psi^z_{2n+1}} \;=\; S( \alpha_{2n+1} , U_{2n+1}, V_{2n+1})\binom{\phi^z_{2n}}{\phi^z_{2n+1}}
\;.
$$
Together with $\CMVR_{1,1} = \langle \phi^z_1, \psi^z_1 \rangle \; = \; U$, one obtains
$$
\CMVR \; = \; U \varoplus \bigoplus_{k\geq 1}\,S(\alpha_{2k+1},U_{2k+1},V_{2k+1}) \;.
$$
Similarly, \eqref{eq-zsego3}  implies
$$
z^{-1} \psi^z_{2n-1} \; =\; \alpha_{2n}\phi^z_{2n-1}  +\rho_{2n} \phi^z_{2n}\;,
\qquad 
z^{-1} \psi^z_{2n} \; = \; (\widetilde{\rho}_{2n})^* \phi^z_{2n-1}  -  V_{2n}\alpha_{2n}^* U_{2n} \phi^z_{2n}. 
$$
Thus, for $n \geq 1$,
$$
 \binom{\psi^z_{2n-1}}{\psi^z_{2n}} \;=\; z\; S( \alpha_{2n} , U_{2n}, V_{2n})\binom{\phi^z_{2n-1}}{\phi^z_{2n}}
 \;, 
$$
and one obtains
$$
\CMVL \; = \; \bigoplus_{k\geq 1}\,S(\alpha_{2k,}U_{2k},V_{2k})\;.
$$
In conclusion, $\CMV$ is a semi-infinite scattering zipper operator.
\hfill $\Box$

\section{Intersection theory and oscillation theorem}

The transfer matrices allow to calculate formal solutions of $\CMV_N\phi^z=z\phi^z$ by equation \eqref{eq-recurrence}. As already pointed out in Section~\ref{sec-transfer}, the $\phi$-component of $\Phi^z$ does not yet lead to a matrix-valued solution of $\CMV_N\phi^z=z\phi^z$ because there is a supplementary constraint on the two components of $\Phi^z_N$. Indeed, because $N$ is even, the last equation of $\CMVR\psi^z=\phi^z$ is $V\phi^z_N=\psi^z_N$. For each vector $v\in\CM^L$ satisfying $\CMVR\psi^zv=\phi^zv$ one finds an eigenvector of $\CMV_N$. Let us reformulate this in terms of the dimension of an intersection of two $L$-dimensional subspaces of $\CM^{2L}$:
\begin{equation}
\label{eq-firstmultiplicity}
\mbox{multiplicity of }z\mbox{ as eigenvalue of }\CMV_N
\;=\;
\dim\bigl( \Phi^z_N\CM^L\cap\Psi_V\CM^L\bigr)
\;,
\end{equation}
where
$$
\Psi_V
\;=\;
\frac{1}{\sqrt{2}}
\;
\begin{pmatrix}
 \one \\ V
\end{pmatrix}
\;.
$$
Now the intersection of the planes $\Phi^z_N\CM^L$ and $\Psi_V\CM^L$ can be conveniently calculated using the fact that both planes are $\Ll$-Lagrangian for $z\in\SM^1$, namely they both satisfy $\Phi^*\Ll\Phi=0$. In fact, the initial condition $\Phi^z_0$ as well as $\Psi_V$ are $\Ll$-Lagrangian as shows a direct calculation. As the transfer matrices $\Tt^z_n$ for $z\in\SM^1$ all conserve the form $\Ll$, also $\Phi^z_n$ is $\Ll$-Lagrangian for all $n$. Before analyzing the intersection of two $\Ll$-Lagrangian planes, let us study the set $\LM_L\subset\GM_L$ of Lagrangian planes, namely those $[\Phi]_\sim\in\GM_L$ satisfying $\Phi^*\Ll\Phi=0$.

\begin{proposi}
\label{prop-Lagrangian} 
$\LM_L\subset\GM_L^\inv$ and $\pi:\LM_L\to\mbox{\rm U}(L)$ is a bijection. Every $[\Phi]_\sim\in\LM_L$ has a representation of the form $\Phi=\binom{U}{\one}$ with $U\in\mbox{\rm U}(L)$.
\end{proposi}

\noindent {\bf Proof.} Let $\Phi=\binom{a}{b}$ be $\Ll$-Lagrangian. Then $\ker(a)\cap\ker(b)=\{0\}$ because otherwise $\Phi$ would not be of rank $L$. Moreover, $a^*a=b^*b$ so that both $a$ and $b$ are invertible. Also $U=ab^{-1}\in\mbox{\rm U}(L)$ and $\Phi=\binom{U}{\one}\,b$. From this all claims follow. 
\hfill $\Box$ 

\vspace{.2cm}

Now follows a general result about the intersection of two $\Ll$-Lagrangian planes.

\begin{proposi}
\label{prop-intersection}  
Let ${\Phi}$ and ${\Psi}$ be $\Ll$-Lagrangian frames and set $W=\pi([\Phi]_\sim)^*\pi([\Psi]_\sim)$. Then
$$
\dim\bigl({\Phi}\,\CM^{L}\,\cap\,{\Psi}\,\CM^{L}\bigr)
\;=\;
\dim\bigl(\mbox{\rm Ker}({\Phi}^*{\Ll}\,{\Psi})\,\bigr)
\;=\;
\mbox{\rm multiplicity of $1$ as eigenvalue of }W\;.
$$
\end{proposi}

\noindent {\bf Proof.} Let us begin with the inequality $\leq$ of the first equality. Suppose there are two $2L\times p$ matrices $v,w$ of rank $p$ such that $\Phi v=\Psi w$. Then $\Phi^*\Ll\Psi w=\Phi^*\Ll\Phi v=0$ so that the kernel of $\Phi^*\Ll\Psi$ is at least of dimension $p$. Inversely, given a $2L\times p$ matrix $w$ of rank $p$ such that $\Phi^*\Ll\Psi w=0$, one deduces that $(\Ll\Phi)^*\Psi w=0$. As the column vectors of $\Phi$ and $\Ll\Phi$ are orthogonal and span $\CM^{L}$, it follows that the column vectors of $\Psi w$ lie in the span of $\Phi$, that is, there exists an $2L\times p$ matrix $v$ of rank $p$ such that $\Psi w=\Phi v$. This shows the other inequality and hence proves the first equality of the lemma. For the second,  let us first note that the dimension of the kernel of $\Phi^*\Ll\Psi$ does not depend on the choice of the representatives. Choosing the representatives $\Phi=\binom{\pi([\Phi]_\sim)}{\one}$ and $\Psi=\binom{\pi([\Psi]_\sim)}{\one}$ then shows the second equality.
\hfill $\Box$

\vspace{.2cm}

For $z\in\SM^1$, let us define $W^z_N\in\mbox{\rm U}(L)$ to be the unitary associated by Proposition~\ref{prop-intersection} to $\Phi=\Phi^z_N$ and $\Psi=\Psi_V$. Taking into account the explicit form of $\Psi_V$ as well as \eqref{eq-Phidef}, one finds
\begin{equation}
\label{eq-UCMVdef} 
W^z_N
\;=\;
\psi_N^z (\phi_N^z)^{-1}V\;,
\end{equation}
These unitaries are the analogs of matrix-valued Pr\"ufer phases and can conveniently be calculated by iterated M\"obius transformations with the transfer matrices. Now the main facts of oscillation theory can be stated.

\begin{theo}
\label{theo-oscillation} 
Let $N\geq 2$ be even. For $z\in\SM^1$, one has
\begin{equation}
\label{eq-eigencalc}
\mbox{\rm multiplicity of $1$ as eigenvalue of }W^z_N
\;=\;
\mbox{\rm multiplicity of $z$ as eigenvalue of }\CMV_N
\;.
\end{equation}
Furthermore, setting $z=e^{\imath\theta}$, 
\begin{equation}
\label{eq-positivity}
\frac{1}{\imath}\,(W^z_N)^*\partial_\theta W^z_N
\;>\;0
\;,
\end{equation}
so that all eigenvalues of $W^z_N$ rotate in the positive sense as a function of $\theta\in\SM^1$.
\end{theo}

\noindent {\bf Proof.}  The equality \eqref{eq-eigencalc} follows immediately from \eqref{eq-firstmultiplicity}, the definition of $W^z_N$ and Proposition~\ref{prop-intersection}. Further, let us note that $W^z_N=\psi_N^z(\phi_N^z)^{-1}V=((\psi_N^z)^*)^{-1}(\phi_N^z)^*V$. Thus one calculates
\begin{eqnarray*}
\frac{1}{\imath}\;
(W^z_N)^*\partial_\theta W^z_N
& = &
\frac{1}{\imath}\;V^*\bigl((\phi_N^z)^{-1}\bigr)^*
\left[(\psi_N^z)^*\partial_\theta\psi_N^z-(\phi_N^z)^*\partial_\theta\phi_N^z\right](\phi_N^z)^{-1}V
\\
& = &
V^*\bigl((\phi_N^z)^{-1}\bigr)^*
(\Phi^z_N)^*(\imath\Ll)(\partial_\theta \Phi^z_N)\,(\phi_N^z)^{-1}V\;.
\end{eqnarray*}
It is therefore sufficient to check the positivity of $(\Phi^z_N)^*(\imath\Ll)\partial_\theta \Phi^z_N$.  By \eqref{eq-recurrence} and $(\Tt^z_n)^*\Ll\Tt^z_n=\Ll$, one now finds
$$
(\Phi^z_N)^*(\imath\Ll)\partial_\theta \Phi^z_N
\;=\;
(\Phi^z_0)^*
\left(
\sum_{n=2}^N
\;
(\Tt^z_{n-1}\cdots\Tt^z_1)^*\,
(\Tt^z_n)^*(\imath\Ll)\partial_\theta \Tt^z_n\,(\Tt^z_{n-1}\cdots\Tt^z_1)\,
\right)
\Phi^z_0
\;.
$$
For odd $n$, the summands vanish because $\Tt^z_n$ is independent of $z$ and thus also $\theta$. We shall now show that for every even $n$ there is a strictly positive definite contribution. For that purpose, let us drop the index $n$ on $\Tt^z_n$ and $S_n$. Now $\Tt^z=\varphi(\overline{z}S)$ is given by
$$
\Tt^z
\;=\;
\begin{pmatrix}
\overline{z} \,A & B \\ C & z\,D
\end{pmatrix}
\;,
$$
where $A$, $B$, $C$ and $D$ are the coefficients of $\Tt^1=\varphi(S)\in\mbox{U}(L,L)$. Then one verifies
$$
(\Tt^z)^*\imath\Ll\,\partial_\theta\Tt^z
\;=\;
\begin{pmatrix}
A^*A & z\, C^*D \\
\overline{z}\,B^*A & D^*D
\end{pmatrix}
\;.
$$
It remains to check that this matrix is positive. On first sight, it is not even hermitian, but actually the defining relations of $\mbox{U}(L,L)$ are 
$$
A^*A-C^*C\;=\;\one\;,
\qquad
D^*D-B^*B\;=\;\one\;,
\qquad
A^*B\;=\;C^*D\;.
$$
The last one shows that the above matrix is indeed hermitian. Now let $\binom{\phi}{\psi}\in\CM^{2L}$ be with, say, $\|\psi\|\geq \|\phi\|$. Then
\begin{eqnarray*}
\binom{\phi}{\psi}^*
\begin{pmatrix}
A^*A & z\, C^*D \\
\overline{z}\,B^*A & D^*D
\end{pmatrix}
\binom{\phi}{\psi}
 & = &
\phi^*A^*A\phi+z\phi^*A^*B\psi+\overline{z}\psi^*B^*A\phi+\psi^*B^*B\psi+\psi^*\psi
\\
& \geq & 
\left(
(\phi^*A^*A\phi)^{\frac{1}{2}}-(\psi^*B^*B\psi)^{\frac{1}{2}}
\right)^2
+\psi^*\psi
\\
& \geq &
\frac{1}{2}\;
\binom{\phi}{\psi}^*\binom{\phi}{\psi}
\;,
\end{eqnarray*}
where in the second step we used the Cauchy-Schwarz inequality. This completes the proof.
\hfill $\Box$

\vspace{.2cm}

\noindent {\bf Remark} As $\CMV_N$ has exactly $NL$ eigenvalues, the rotation number of $z\in\SM^1\mapsto W^z_N$ is equal to $NL$. This can also be shown independently by calculating the Maslov index as in \cite{SB1}.
\hfill $\diamond$

\section{Oscillation theory for finite periodic scattering zipper}
\label{sec-periodic}

It is possible to associate to a sequence $(S_n)_{n=1,\ldots,N}$ of scattering matrices $S_n\in\mbox{\rm U}(2L)_\inv$ with $N$ still even, a periodic scattering zipper operator $\CMV_N^\per=\CMVL^\per_N\CMVR_N^\per$ by setting
$$
\CMVL^\per_N
\;=\; 
\begin{pmatrix}
S_2 & & & & \\
 & S_4 & & & \\
& & \ddots & & \\
& & & \ddots & \\
& & & & S_{N}
\end{pmatrix}                                                                                            
\;,
\qquad
\CMVR_N^\per\;=\;
\begin{pmatrix}
\delta_1 & & & & \gamma_1 \\
& S_3  & & & \\
 & & \ddots  & & \\
& & & S_{N-1}  & \\
\beta_1 & & & & \alpha_1
\end{pmatrix}                                                                                            
\;,
$$
where
$$
S_1\;=\;
\begin{pmatrix}
\alpha_1 & \beta_1 \\
\gamma_1 & \delta_1 
\end{pmatrix}
\;.
$$
The aim of this section is to calculate the spectrum of $\CMV_N^\per$. This parallels the calculations in \cite{SB2} and is useful for the calculation of the spectrum of infinite periodic scattering zippers, as explained in Section~\ref{sec-periodic-infinite}. The solutions $\phi^z\in\ell^2(\{1,\ldots,N\})\otimes\CM^L$ of the eigenvalue equation  $\CMV_N^\per\phi^z=z\phi^z$ can again be constructed with the transfer matrices. Indeed, every eigenvector of $\Tt^z(N,0)$ to the eigenvalue $1$ allows to construct a periodic eigenvector so that
$$
\mbox{\rm multiplicity of }z\;\mbox{\rm as eigenvalue of }\CMV_N^\per
\;=\;
\mbox{\rm multiplicity of }1\;\mbox{\rm as eigenvalue of }\Tt^z (N,0)
\;.
$$
Hence one needs to find those $z\in\SM^1$ for which $1$ is eigenvalue of $\Tt^z (N,0)$. By considering the fact that the graph of $\Tt^z (N,0)$ is also Lagrangian w.r.t. an adequate quadratic form, this can be done by means of intersection theory in a similar manner as in the previous section. 

\vspace{.2cm}

Several notations need to be introduced. Let us associate to $\Tt$ the $4L\times 4L$ matrix $\widehat{\Tt}=\one_{2L}\widehat{\oplus}\Tt$ where the $\widehat{\oplus}$ denotes the checker board sum given by
\begin{equation}
\label{eq-expanding}
\left(
\begin{array}{cc}
A & B \\
C & D
\end{array}
\right)
\,\widehat{\oplus}\,
\left(
\begin{array}{cc}
A' & B' \\
C' & D'
\end{array}
\right)
\;=\;
\left(
\begin{array}{cccc}
A & 0 & B & 0 \\
0 & A' & 0 & B' \\
C & 0 & D & 0 \\
0 & C' & 0 & D'
\end{array}
\right)
\;.
\end{equation}
This allows to define the quadratic form $\widehat{\Ll}={\Ll}\,\widehat{\oplus}\,\Ll$. For $z\in\SM^1$ the matrices $\widehat{\mathcal{T}}^z_n$ conserve the quadratic form $\widehat{\Ll}$. Therefore also $\widehat{\Ll}$-Lagrangian planes are mapped onto $\widehat{\Ll}$-Lagrangian planes. By Proposition~\ref{prop-Lagrangian}, the stereographic projection maps the set $\LM_{2L}$ of $\widehat{\Ll}$-Lagrangian planes is diffeomophically to $\mbox{\rm U}(2L)$. Let us denote the stereographic projection by $\widehat{\pi}:\LM_{2L}\to\mbox{\rm U}(2L)$. 

\begin{proposi}
\label{prop-TtoU0} 
To a given $\Tt\in\mbox{\rm U}(L,L)$, let us associate the unitary
\begin{equation}
\label{eq-Uhatdef}
\widehat{W}
\;=\;
\widehat{\pi}([\widehat{\Tt}\,\widehat{\Psi}_0]_\sim)^*\;
\widehat{\pi}([\widehat{\Psi}_0]_\sim)
\;\in\;
\mbox{\rm U}(2L)
\;,
\end{equation} 
where the $\widehat{\Ll}$-Lagrangian plane is given by the $4L\times 2L$ matrix
\begin{equation}
\label{eq-Phiexpand}
\widehat{\Psi}_0
\;=\;
\left(
\begin{array}{cc}
0 & \one \\
\one & 0 \\
\one & 0  \\ 
0 & \one \end{array}
\right)
\;.
\end{equation}
Then
$$
\mbox{\rm geometric multiplicity of }1\;\mbox{\rm as eigenvalue of }\Tt
\;=\;
\mbox{\rm multiplicity of }1\;\mbox{\rm as eigenvalue of }\widehat{W}\;.
$$
\end{proposi}

\noindent {\bf Proof.} First of all, it can readily be checked that $\widehat{\Psi}_0^*\widehat{\Ll}\,\widehat{\Psi}_0=0$.  Now let us suppose that this frame and the Lagrangian frame $\widehat{\Tt}\,\widehat{\Psi}_0$ have a non-trivial intersection. This means that there exist vectors $v,w,v',w'\in\CM^L$ such that such $\widehat{\Psi}_0\binom{v}{w}=\widehat{\Tt}\,\widehat{\Psi}_0\binom{v'}{w'}$. The first and third line of this vector equality imply $w=w'$ and $v=v'$, the other two that $\Tt\binom{v}{w}=\binom{v}{w}$. This shows 
$$
\mbox{\rm geometric multiplicity of }1\;\mbox{\rm as eigenvalue of }\Tt
\;=\;
\dim\bigl(\,\widehat{\Tt}\,\widehat{\Psi}_0\,\CM^{2L}\,\cap\,
\widehat{\Psi}_0\,\CM^{2L}\bigr)
\;.
$$
But now Proposition~\ref{prop-intersection} can be applied to calculate the r.h.s. and completes the proof.
\hfill $\Box$

\vspace{.2cm}
 
It is now natural to introduce the following $\widehat{\Ll}$-Lagrangian frames:
\begin{equation}
\label{eq-transferidexpand}
\widehat{\Psi}^z_n
\;=\;
\widehat{\Tt}^z_n
\;\widehat{\Psi}^z_{n-1}
\;,
\qquad
\widehat{\Psi}^z_0
\;=\;
\widehat{\Psi}_0
\;,
\qquad
n \geq 1,
\end{equation}
with $\widehat{\Psi}_0$ as in \eqref{eq-Phiexpand}.  Associated are then the unitaries
$$
\widehat{W}^z_N
\;=\;
\widehat{\pi}\bigl([ \widehat{\Psi}^z_N]_\sim\bigr)^*\;
\widehat{\pi}\bigl([ \widehat{\Psi}_0]_\sim\bigr)
\;.
$$
%

\begin{theo} 
\label{theo-osci2} 
The multiplicity of $z = e^{i\theta} \in \SM^1$ as eigenvalues of $\CMV_N^\per$ is equal to the multiplicity of $1$ as eigenvalue of $\widehat{W}^z_N$. Moreover,
$$
\frac{1}{\imath}\;(\widehat{W}^{z}_N)^*\,\partial_\theta\,\widehat{W}^{z}_N \; > \; 0
\;,
$$
so that the eigenvalues of $\widehat{W}^{z}_N$ rotate around the unit circle in the positive sense and with non-vanishing speed as function of $\theta$.
\end{theo}

\noindent {\bf Proof.} The first claim follows directly from Proposition~\ref{prop-TtoU0}. For the proof  of the second one, let us denote the upper and lower entries of $\widehat{\Psi}^z_N$ by $\psi^z_+$ and $\psi^z_-$. These are $2L\times 2L$ matrices such that $\widehat{\pi}\bigl([ \widehat{\Psi}^z_n]_\sim\bigr)=\psi_-^z(\psi_+^z)^{-1}=((\psi_-^z)^{-1})^*(\psi_+^z)^*$. Then
$$
(\widehat{W}^{z}_N)^*\,\partial_\theta\,\widehat{W}^{z}_N
\;=\;
\widehat{\pi}\bigl([ \widehat{\Psi}_0]_\sim\bigr)^*\;
((\psi_+^z)^{-1})^*
\Bigl[\,
(\psi_+^z)^*\partial_\theta\psi_+^z \,-\,
(\psi_-^z)^*\partial_\theta\psi_-^z
\,\Bigr]
(\psi_+^z)^{-1}
\;\widehat{\pi}\bigl([ \widehat{\Psi}_0]_\sim\bigr)
\;.
$$
Thus it is sufficient to check positive definiteness of
$$
\imath\;
\Bigl[\,
(\psi_-^z)^*\partial_\theta\psi_-^z \,-\,
(\psi_+^z)^*\partial_\theta\psi_+^z
\,\Bigr]
\;=\; \imath \;
(\widehat{\Psi}^z_N)^*\,\widehat{\Ll}\,\partial_\theta \widehat{\Psi}^z_N
\;.
$$
From the product rule follows that
$$
\partial_\theta \widehat{\Psi}^z_N
\;=\;
\sum_{n=1}^N
\;
\left(
\prod_{l=n+1}^N\,\widehat{\Tt}^z_l
\right)
\,
\left(\partial_\theta \widehat{\Tt}^z_n\right)
\;
\left(
\prod_{l=1}^{n-1}\,\widehat{\Tt}^z_l
\right)
\,\widehat{\Psi}_0^z
\;.
$$
This implies that
$$
\imath \; (\widehat{\Psi}^z_N)^*\,\widehat{\Ll}\,\partial_\theta \widehat{\Psi}^z_N
\;=\;
\sum_{n=1}^N
\;(\widehat{\Psi}_0^z)^*\,
\left(
\prod_{l=1}^{n-1}\,\widehat{\Tt}^z_l
\right)^*
\,
\bigl(\widehat{\Tt}^z_n\bigr)^*\,\imath\; \widehat{\Ll}\,
\bigl(\partial_\theta \widehat{\Tt}^z_n\bigr)
\;
\left(
\prod_{l=1}^{n-1}\,\widehat{\Tt}^z_l
\right)
\,\widehat{\Psi}_0^z
\;.
$$
As
$$
\bigl(\widehat{\Tt}^z_n\bigr)^*\, (\imath\;\widehat{\Ll})\,
\bigl(\partial_\theta \widehat{\Tt}^z_n\bigr)
\;=\; 
{\bf 0}\; \widehat{\varoplus}\; \bigl((\Tt^z_n)^* \, (\imath \, \Ll) \,\partial_\theta \Tt^z_n\bigr)
\;,
$$
and the matrices $\widehat{\Tt}^E_n$ do not mix first and third columns and lines with the rest, it follows from evaluation in the state $\widehat{\Psi}_0$ that
$$
(\widehat{\Psi}^z_N)^*\,\imath \: \widehat{\Ll}\,\partial_\theta \widehat{\Psi}^z_N
\;=\;
\sum_{n=1}^N
\;
\left(
\prod_{l=1}^{n-1}\,{\Tt}^z_l
\right)^*
\,
\bigl((\Tt^z_n)^* \, (\imath \, \Ll) \,\partial_\theta \Tt^z_n\bigr)
\;
\left(
\prod_{l=1}^{n-1}\,{\Tt}^z_l
\right)
\;.
$$
But positivity of $(\Tt^z_n)^* \, (\imath \, \Ll) \,\partial_\theta \Tt^z_n$ was already checked in the proof of Theorem~\ref{theo-oscillation}. This completes the proof.
\hfill $\Box$

\section{Spectrum of infinite periodic scattering zippers} 
\label{sec-periodic-infinite}

In this section, we consider a two-sided infinite scattering zipper $\CMV^{\mbox{\tiny per}}$ defined on $\ell^2(\ZM,\CM^L)$ which are $N$-periodic where again $N$ is even. It is specified by a sequence of scattering matrices $S_n\in\mbox{U}(2L)_\inv$ satisfying $S_n = S_{n+N}$ for all $n\in\ZM$. One can partially diagonalize such periodic operators by the Bloch-Floquet transform defined next. 

\begin{defini}
The Bloch-Floquet transform $\Ff:\ell^2(\mathbb{Z})\varotimes \CM^{L}\to L^2(\TM_N) \varotimes \CM^N \varotimes \CM^{L}$ is defined by
$$
(\Ff \phi)_n (k) \; = \; \frac{1}{\sqrt{|\TM_N|}} \sum_{m \in \ZM} \phi_{n+mN} \;e^{\imath(n+mN)k}, \qquad n \in {0,\dots, N-1}\,, \;\; k \in \TM_N\,,
$$  
where $\TM_N=(-\frac{\pi}{N},\frac{\pi}{N}]$ and $|\TM_N|=\frac{2\pi}{N}$. Its inverse $\Ff^{-1}:L^2(\TM_N) \varotimes \CM^N \varotimes \CM^{L}\to\ell^2(\mathbb{Z})\varotimes \CM^{L}$ is given by
$$
(\Ff^{-1} \phi)_n \; = \; \frac{1}{\sqrt{|\TM_N|}} \int_{\TM_N} dk \; \phi_{n \; \mbox{\tiny\rm mod} \, N} (k) \; e^{-\imath kn}\,, \qquad n \in \ZM\,. 
$$
\end{defini} 

\begin{proposi}\label{prop-fourier}
One has the following properties:

\vspace{.1cm}

\noindent {\rm (i)} $\Ff^{-1} \, = \, \Ff^*$, namely $\Ff$ is unitary.

\vspace{.1cm}

\noindent {\rm (ii)} Let $T$ be the shift on $\ell^2(\ZM)\varotimes \CM^{L}$ defined by $T\phi_n = \phi_{n+1}$ and let $T_{\mbox{\tiny\rm cyc}}$ be the cyclic shift on $\CM^N$. 

Then
$$
(\Ff \,T \Ff^* \phi)_n(k) \;= \;e^{-ik}\, (T_{\mbox {\tiny \rm cyc}} \phi)_n(k)\,, 
\qquad n \in \ZM\,, \;\; k \in \TM_N\,.
$$
\end{proposi}

\noindent{\bf Proof.}
This follows from direct computations. 
\hfill $\Box$

\vspace{2mm}

For any $k \in \TM_N$ and $S_j \, = \,S(\alpha_j,U_j,V_j) \in \mbox{U}(2L)_\inv$ let us set 
$$
 S_j(k) \;=\; 
\begin{pmatrix} \alpha_j & e^{-\imath k} \,(\one-\alpha_j\alpha_j^*)^{\frac{1}{2}}U_j  
\\ 
e^{\imath k}\,V_j(\one-\alpha_j^*\alpha_j)^{\frac{1}{2}} & -V_j\alpha_j^*U_j 
\end{pmatrix}
\;.
$$
Now $\CMVL^{ \mbox{\tiny \rm per}}_N(k)$ and $\CMVR^{ \mbox{\tiny\rm per}}_N(k)$ are defined as Section~\ref{sec-periodic} using $S_j(k)$ instead of $S_j$ and then $\CMV^{ \mbox{\tiny\rm per}}_N(k) = \CMVL^{ \mbox{\tiny\rm per}}_N(k)\CMVR^{ \mbox{\tiny\rm per}}_N(k)$ is a finite periodic scattering zipper, the spectrum of which can be calculated by the technique of Section~\ref{sec-periodic}. By the following result this allows to calculate the spectrum of $\CMV^{\mbox{\tiny per}}$.

\begin{theo}
\label{theo-oscillationper}
The operators $\CMVL^{ \mbox{\tiny\rm per}}$, $\CMVR^{ \mbox{\tiny\rm per}}$ and $\CMV^{ \mbox{\tiny\rm per}}$ are fibered after Bloch-Floquet transformation, precisely
$$
\Ff\, \CMVL^{\mbox{\tiny\rm per}} \Ff^* \,=\,  \int_{\TM_N}^\oplus dk \, \CMVL^{ \mbox{\tiny\rm per}}_N(k)\,, 
\qquad  \Ff \,\CMVR^{\mbox{\tiny\rm per}} \Ff^*  \,=\,  \int_{\TM_N}^\oplus dk \, \CMVR^{ \mbox{\tiny\rm per}}_N(k)\,, 
\qquad \Ff \,\CMV^{\mbox{\tiny\rm per}} \Ff^*  \,=\,  \int_{\TM_N}^\oplus dk \, \CMV^{ \mbox{\tiny\rm per}}_N(k)\,.
$$
Therefore,
$$
\sigma (\CMV^{\mbox{\tiny\rm per}}) 
\; = \; \bigcup_{k \,\in\, \TM_N} \sigma \left( \CMV^{ \mbox{\tiny\rm per}}_N(k)\right)
\;.
$$
\end{theo}

\noindent {\bf Proof.}
By definition, for $\phi \in L^2(\TM_N)\varotimes \CM^N \varotimes \CM^{L}$:
$$
(\Ff\, \CMVL^{\mbox{\tiny\rm per}} \Ff^* \phi)_n (k) \;=\; \frac{1}{\sqrt{|\TM_N|}} \,
\sum_{m \in \ZM}\, ( \CMVL^{\mbox{\tiny\rm per}}\Ff^* \phi)_{n+mN} \; e^{\imath(n+mN)k}
\;.
$$
In case $n=2j-1$, one obtains from the structure of $\CMVL^{\mbox{\tiny\rm per}}$:
\begin{eqnarray*}
& & 
\!\!\!\!\!\!\!\!\!\!\!\!\!\!
(\Ff\, \CMVL^{\mbox{\tiny\rm per}} \Ff^*\phi)_{2j-1} (k) \\
& & =\;\frac{1}{\sqrt{|\TM_N|}} \,
\sum_{m \in \ZM} \left( \alpha_{2j} (\Ff^* \phi)_{2j-1+mN} + (\one-\alpha_{2j}\alpha_{2j}^*)^{\frac{1}{2}}U_{2j} (\Ff^* \phi)_{2j+mN} \right) e^{\imath(2j-1+mN)k} \\
& & = \; \alpha_{2j} \; \phi_{2j-1} \;+\; (\one-\alpha_{2j}\alpha_{2j}^*)^{\frac{1}{2}}U_{2j} \;e^{-\imath k} \;\phi_{2j}\;.
\end{eqnarray*}
By the same calculation for $n=2j$,
$$
(\Ff\, \CMVL^{\mbox{\tiny\rm per}} \Ff^* \phi)_{2j} (k) \;=\; 
V_{2j}(\one-\alpha_{2j}^*\alpha_{2j})^{\frac{1}{2}}e^{\imath k} \; \phi_{2j-1} \;  -\; V_{2j}\alpha_{2j}^*U_{2j} \;\phi_{2j}
\;.
$$
Together this shows 
$$
\Ff\, \CMVL^{\mbox{\tiny\rm per}} \Ff^* \;=\; \int_{\TM_N}^\oplus dk \; S_2(k) \varoplus S_4(k) \varoplus \dots \varoplus S_N(k) \;=\; \int_{\TM_N}^\oplus dk \; \CMVL^{\mbox{\tiny\rm per}}_N(k)
\;.
$$
\noindent The second equality is proved using Proposition~\ref{prop-fourier}(ii) and previous computation:
\begin{align*}
\Ff \,\CMVR^{\mbox{\tiny\rm per}} \Ff^* \;&=\; (\Ff\, T \Ff^*)(\Ff \,T^{-1}\CMVR^{\mbox{\tiny\rm per}}T \Ff^* )(\Ff \,T^{-1} \Ff^*) \\
&= \; \int_{\TM_N}^\oplus dk \; T_{\mbox {\tiny\rm cyc}} \left( S_1(k) \varoplus \dots \varoplus S_{N-1}(k)\right) T_{\mbox {\tiny\rm cyc}}^{-1} \; = \; \int_{\TM_N}^\oplus dk \; \CMVR^{\mbox{\tiny\rm per}}_N(k)
\;.
\end{align*} 
As the product of fibered operators is fibered, this also implies the formula for $\CMV^{\mbox{\tiny\rm per}}$.
\hfill  $\Box$

\section*{Appendix A: M\"obius transformations}
\label{sec-Moeb}

This appendix resembles some basic properties of the M\"obius transformation as
they are used in the main text. A lot of references to the literature can be
found in \cite{DPS}. The M\"obius transformation (also called canonical
transformation or fractional transformation) is defined by
\begin{equation}
\label{eq-moebius}
\Tt\cdot Z
\;=\;
(AZ+B)\,(CZ+D)^{-1}
\,,
\qquad
\Tt
\,=\,
\left(
\begin{array}{cc} A & B \\  C & D \end{array}
\right)
\;\in\;\mbox{Gl}(2L,\CM)
\,,
\;\;Z\in\mbox{Mat}(L\times L,\CM)\,,
\end{equation}
whenever the appearing inverse exists. If $\Tt$ is $\Jj$-unitary and
$Z\in\HM_L$, then $\Tt\cdot Z$ exists and is in $\HM_L$ (see Appendix~B for the definition of the upper half-plane $\HM_L$). For $\Tt$ as in
\eqref{eq-moebius} and as long as the appearing inverse exists, the inverse
M\"obius transformation is defined by
\begin{equation}
\label{eq-moebiusinv} W:\Tt \;=\; (WC-A)^{-1}\,(B-WD) \,, \qquad
W\in\mbox{Mat}(L\times L,\CM)\,.
\end{equation}
The M\"obius transformation is a left action, namely $(\Tt\Tt')\cdot
Z=\Tt\cdot(\Tt'\cdot Z)$ as long as all objects are well-defined. The inverse
M\"obius transformation is a right action in the sense of the following
proposition, the algebraic proof of which is left to the reader.

\begin{proposi}
\label{prop-Moebinv} Under the condition that all the M\"obius und inverse
M\"obius transformations as well as matrix inverses below exist, one has the
following properties.

\vspace{.2cm}

\noindent {\rm (i)} $\;\;W=\Tt\cdot Z\;\;\Leftrightarrow\;\; W:\Tt=Z$

\vspace{.1cm}

\noindent {\rm (ii)} $\;W:(\Tt\Tt')=(W:\Tt):\Tt'$

\vspace{.1cm}

\noindent {\rm (iii)} $W:\Tt=\Tt^{-1}\cdot W$

\end{proposi}

\section*{Appendix B: Riesz-Herglotz representation theorem}
\label{sec-positivity}

Let $\HM_L$ denote the upper half plane of matrices $Z\in\mbox{\rm Mat}(L,\CM)$ such that $\Im m(Z)=\imath(Z^*-Z)>0$. It is well-known ({\it e.g.} Section 4.5. of \cite{Sim}) that the Cayley transform maps $\HM_L$ via M\"obius transformation to the Siegel disc $\DM_L$, namely $\Cc\cdot \HM_L=\DM_L$.  An analytic function $z\in\HM_1\mapsto G(z)\in\HM_L$ is called a Herglotz function. Then $F(z)=G(\Cc^*\cdot z)$ is an analytic function on the unit disc $\DM_1$ having positive imaginary part. If, moreover, $F(0)=\imath\,\one$, then such a function is called a Caratheodory function. The scalar version of the following classical theorem can be found in text books such as \cite{Lax}. The matrix version is an immediate corollary of it.

\begin{theo} {\rm (Riesz-Herglotz representation theorem)}
Let $F:\DM\to \mbox{\rm Mat}(L\times L,\CM)$ be analytic satisfying $\Im m(F(z))>0$ and $F(0)=\imath\,\one$. Then there exists a unique matrix-valued probability measure $\mu$ on $\SM^1$ such that
$$
F(z)
\;=\;
\imath\,\int_{\SM^1} \mu(d\xi)\;\frac{\xi+z}{\xi-z}
\;.
$$
\end{theo}

\end{document}